\documentclass[pdflatex,sn-mathphys-num]{sn-jnl}

\usepackage{graphicx}%
\usepackage{multirow}%
\usepackage{amsmath,amssymb,amsfonts}%
\usepackage{amsthm}%
\usepackage{mathrsfs}%
\usepackage[title]{appendix}%
\usepackage{xcolor}%
\usepackage{textcomp}%
\usepackage{manyfoot}%
\usepackage{booktabs}%
\usepackage{algorithm}%
\usepackage{algorithmicx}%
\usepackage{algpseudocode}%
\usepackage{listings}%
\usepackage{graphicx}
\usepackage{subcaption}
\usepackage{comment}
\usepackage[export]{adjustbox}
\usepackage{float}
\usepackage{placeins}
\usepackage{color,soul}
\usepackage{amssymb}

\theoremstyle{thmstyleone}%
\theoremstyle{thmstyletwo}%

\theoremstyle{thmstylethree}%

\begin{document}

\title[Unsupervised Multimodal Graph-based Model for Geo-social Analysis]{Unsupervised Multimodal Graph-based Model for Geo-social Analysis}


\author[1]{\fnm{Ehsaneddin} \sur{Jalilian}}\email{ehsaneddin.jalilian@it-u.at}

\author[1,2]{\fnm{Bernd} \sur{Resch}}\email{bernd.resch@it-u.at}


\affil[1]{\orgdiv{GeoSocial Artificial Intelligence}, \orgname{Interdisciplinary Transformation University Austria}, \orgaddress{\street{Altenberger Straße 66c}, \city{Linz}, \postcode{A-4040}, \state{Upper Austria}, \country{Austria}}}

\affil[2]{\orgdiv{Center for Geographic Analysis}, \orgname{Harvard University}, \orgaddress{\street{1737 Cambridge Street}, \city{Cambridge}, \postcode{02138}, \state{MA}, \country{USA}}}



\abstract{The systematic analysis of user-generated social media content, especially when enriched with geospatial context, plays a vital role in domains such as disaster management and public opinion monitoring. Although multimodal approaches have made significant progress, most existing models remain fragmented, processing each modality separately rather than integrating them into a unified end-to-end model. To address this, we propose an unsupervised, multimodal graph-based methodology that jointly embeds semantic and geographic information into a shared representation space. The proposed methodology comprises two architectural paradigms: a mono graph (MonoGrah) model that jointly encodes both modalities, and a multi graph (MultiGraph) model that separately models semantic and geographic relationships and subsequently integrates them through multi-head attention mechanisms. A composite loss, combining contrastive, coherence, and alignment objectives, guides the learning process to produce semantically coherent and spatially compact clusters. Experiments on four real-world disaster datasets demonstrate that our models consistently outperform existing baselines in topic quality, spatial coherence, and interpretability. Inherently domain-independent, the framework can be readily extended to diverse forms of multimodal data and a wide range of downstream analysis tasks.}




\maketitle

\section{Introduction}\label{sec1}

Social media platforms continuously generate vast volumes of user-generated content, including posts, comments, images, videos, and interaction signals such as likes and shares. This content is rich, diverse, and real-time, offering a valuable resource for understanding how people communicate, express opinions, form communities, and respond to events. The widespread adoption of smartphones equipped with location-tracking capabilities has added an important additional dimension to this landscape: geographic context. By enabling users to share their locations, these devices have facilitated the integration of geospatial information into online content, significantly enhancing the contextual depth of social media analysis.

A growing body of research has demonstrated the value of geo-referenced social media data across a wide range of applications, including disaster response~\cite{goodchild2010crowdsourcing}, public health monitoring~\cite{Signorini2011}, and crisis management. In particular, location-aware content from microblogging platforms has been leveraged to study and monitor phenomena such as earthquakes~\cite{Havas2021Portability}, floods~\cite{Huang2018Flood}, refugee movements~\cite{Havas2021Refugee}, and disease outbreaks~\cite{Stolerman2023Disease}. These studies highlight the potential of combining textual and spatial signals to gain timely and actionable insights from social media streams.

Much of the existing social media analysis literature has focused on semantic topic modeling as a fundamental technique for uncovering latent thematic structures within large collections of user-generated text~\cite{Resch2018,Hannigan2019}. While effective in capturing semantic regularities, these approaches typically treat text in isolation and rarely integrate additional modalities in a principled manner. Early efforts toward multimodal geo-social analysis often relied on sequential or modular workflows, in which each modality (e.g., text, location, time) was processed independently before being combined at a later stage~\cite{yin2011geographical}. Such designs limit the ability to capture complex cross-modal interactions and often introduce brittle dependencies between processing stages, making results sensitive to preprocessing choices and execution order.

More recent approaches have advanced toward integrating multiple modalities, including temporal signals, geographic space, semantic topics, and sentiment~\cite{gonzalez2024neural,hanny2025}. However, these methods still depend on fragmented pipelines with multiple isolated components, falling short of truly end-to-end multimodal learning. The inherent heterogeneity of input modalities further complicates this problem: text is symbolic and sequential, while geographic data is continuous and spatial. As a result, feature representations across modalities are often misaligned, leading to inconsistencies during fusion and reduced interpretability. The absence of a unified learning framework constrains cross-modal reasoning and limits the model’s ability to jointly exploit semantic and spatial structure.

In this paper, we propose a unified multimodal graph-based learning framework that jointly embeds semantic and geographic information into a shared representation space using end-to-end training. Our methodology constructs parallel semantic and geographic graphs over social media posts and learns representations that are both semantically coherent and spatially interpretable. Nodes in the graph correspond to post-level embeddings, while edges encode relational signals such as semantic similarity and geographic proximity. This formulation enables the model to reason over multimodal interactions in a structured and context-aware manner, rather than relying on post hoc fusion of independently learned features.

At the core of our approach is an unsupervised graph neural network that aggregates information from each node’s neighborhood through a weighted message-passing mechanism. A key innovation of the proposed framework is the use of a composite loss function inspired by reference-less clustering evaluation principles. The loss integrates three complementary objectives: a contrastive loss that preserves local semantic neighborhoods by pulling similar nodes closer while pushing dissimilar ones apart; a coherence loss that promotes compact and well-separated clusters; and an alignment loss that stabilizes training by encouraging consistency between node embeddings and their assigned cluster centroids. By embedding these objectives directly into the learning process, the model discovers topic clusters that are simultaneously semantically meaningful and geographically grounded.

Our approach is motivated by applications in disaster management, where geo-social media provides rich, fine-grained information from affected communities during emergencies. Through a case study of the 2017 Hurricane Harvey event in the United States, we demonstrate that the proposed framework produces geographically localized and semantically coherent topic clusters that closely align with observed disaster impacts. While this case study illustrates the practical utility of the model in crisis scenarios, the framework itself is application-agnostic and can be readily adapted to other multimodal data integration tasks across domains such as public health, urban analytics, and social sensing.

In summary, this work makes the following contributions:
\begin{itemize}
    \item We introduce a unified multimodal graph-based framework that integrates semantic and geospatial information into a shared representation space for unsupervised topic discovery.
    \item We propose two unsupervised architectures, MonoGraph and MultiGraph, that flexibly support different levels of multimodal interaction while remaining domain-agnostic.
    \item We develop a composite loss function combining contrastive, coherence, and alignment objectives to generate clusters that are both semantically coherent and spatially interpretable.
    \item We validate the effectiveness of the proposed approach through a real-world disaster case study, demonstrating improved topic quality, spatial compactness, and interpretability compared to existing baselines.
\end{itemize}

\section{Related work}\label{Sec-2}

To contextualize our work, we first examine general topic modeling techniques and their geospatial adaptations, specifically tailored for analyzing social media data. We then delve into graph-based topic modeling approaches, emphasizing their strengths in capturing intricate relationships within the data.

\subsection{Topic Modeling}\label{sec2-1}

Topic modeling is a widely used technique for discovering latent semantic topics within large document collections, providing a structured thematic summary by assigning one or more topic labels to each document~\cite{hoyle2022content}. Among the foundational methods, Latent Dirichlet Allocation (LDA)~\cite{blei2003latent} introduced a fully generative probabilistic framework in which each document is modeled as a mixture over latent topics, and each topic is represented as a multinomial distribution over words. The LDA framework has been extended to incorporate additional dimensions such as sentiment~\cite{Lin2009Joint}, temporal dynamics~\cite{Blei2006Dynamic}, and user-level information in social media~\cite{zhao2011comparing}, broadening the applicability of topic models across domains. However, LDA and its variants fundamentally rely on the bag-of-words assumption, which ignores word order and semantic relationships between words, limiting their effectiveness in handling short, noisy, or highly contextual texts such as tweets and social media posts~\cite{hong2010empirical}. To overcome these limitations, embedding-based topic modeling techniques have recently emerged. Dieng et al.~\cite{dieng2020topic} introduced the Embedded Topic Model (ETM), which integrates pre-trained word and sentence embeddings with LDA’s generative framework to better capture semantic relationships. Sia et al.~\cite{sia2020neural} further demonstrated that clustering high-dimensional semantic embedding vectors, derived from neural language models, can serve as a powerful alternative strategy, achieving performance comparable to traditional topic models. 

Building on these embedding-based approaches, Angelov~\cite{angelov2020top2vec} proposed Top2Vec, which leverages document embeddings and clustering algorithms to identify topics without requiring the number of topics to be pre-specified. This method utilizes techniques like Uniform Manifold Approximation and Projection (UMAP) and Hierarchical Density-Based Spatial Clustering of Applications with Noise (HDBSCAN) for dimensionality reduction and flexible clustering. Grootendorst~\cite{grootendorst2022bertopic} further advanced this line of work with BERTopic, which combines transformer-based contextual embeddings with refined clustering strategies, consistently outperforming traditional LDA and Top2Vec in both coherence and interpretability. Complementary empirical evidence from related domains further highlights the limitations of purely embedding-driven approaches. In particular, studies of pre-trained language models in simple knowledge graph question answering (KGQA) show that, while contextual embeddings capture strong lexical and semantic regularities, they struggle to reliably encode explicit relational structure without dedicated graph-based inductive bias~\cite{hu2023empirical}. These findings suggest that high-quality semantic embeddings alone are often insufficient for tasks requiring structured relational reasoning, motivating approaches that integrate neural language representations with explicit graph modeling.

\subsection{Spatio-semantic Topic Modeling}\label{sec2-2}

As location-aware social media platforms proliferate, spatially grounded topic modeling has gained increasing attention. This line of work focuses on uncovering geographically coherent semantic patterns by integrating geospatial signals into the semantic topic modeling process. Early efforts extended classical models like LDA with spatial priors. For example, Spatial LDA (sLDA)~\cite{Yuan2013Who} and Geo-aware Topic Models~\cite{Hong2012Discovering} include location-based information as latent variables or observed metadata, enabling the discovery of region-specific topics. Similar models have been applied to a variety of domains, such as understanding natural disasters~\cite{Resch2018}, flood mapping~\cite{Huang2018Flood}, and epidemic monitoring~\cite{Stolerman2023Disease}. However, most spatial topic models treat location and text independently or as loosely coupled variables, which often leads to weak cross-modal alignment. In addition, spatial topic models are frequently built upon modular architectures, making them difficult to scale or generalize. More recent multimodal approaches attempt to combine geolocation, text, and other modalities, often through hybrid processing pipelines~\cite{gonzalez2024neural}~\cite{hanny2025}. While these methods capture richer context, they extract each modality feature in isolation and then merge features post hoc, limiting their ability to jointly reason across modalities. The absence of a unified representation space, together with the gap between symbolic textual data and continuous geospatial representations, further introduces alignment and interpretability challenges in downstream clustering or topic interpretation.

\subsection{Graph-based Topic Modeling}\label{sec2.3}

Graph-based topic modeling has emerged as an effective alternative to traditional probabilistic models by explicitly exploiting relational structure to improve semantic coherence, particularly for short and sparse texts. Early approaches construct word co-occurrence graphs from text corpora and apply community detection algorithms to identify topical clusters. For instance, Wang et al.~\cite{wang2015community} demonstrated that community detection methods such as Louvain can outperform Latent Dirichlet Allocation (LDA) on short text datasets. Similarly, Zhao et al.~\cite{zhao2018graph} proposed the Graph-based Short Text Topic Model (GSTM), which leverages word co-occurrence graphs to alleviate data sparsity and improve topic coherence.

Subsequent work has sought to integrate graph structure more deeply into topic modeling. Newman and Girvan~\cite{newman2006finding} established a theoretical connection between community detection in bipartite document--word networks and topic model inference, highlighting the natural synergy between network science and probabilistic topic discovery. Extensions of LDA incorporating graph structure, such as Graph-LDA~\cite{doshi2015}, embed word similarity relations directly into the generative framework, yielding more semantically coherent topics. More recently, Graph Topic Models (GTMs) have combined Graph Neural Networks (GNNs) with document--word heterogeneous graphs to jointly learn semantic representations and relational structure~\cite{zhou2020}. While these models benefit from relational inductive biases, they often struggle to capture deeper or long-range semantic dependencies and remain limited in modeling more abstract or nuanced topics.

Embedding-based graph clustering approaches further extend graph-based topic modeling by operating directly on learned sentence or document representations. Graph2Topic~\cite{graph2topic2023}, for example, applies graph clustering to embedding-based representations, improving topic interpretability and quality without requiring a predefined number of topics. This line of work demonstrates the practical advantages of combining graph structure with embedding-based semantic representations for topic discovery. In parallel, recent advances in generic graph representation learning have introduced self-supervised contrastive objectives for learning node or graph embeddings without labels. GraphCL~\cite{you2020graphcl} performs graph-level contrastive learning by contrasting different augmented views of the same graph, while MVGRL~\cite{hassani2020} learns node representations by contrasting multiple structural views of a graph. Although these methods achieve strong performance on generic graph learning benchmarks, they are not designed for topic discovery and do not explicitly model semantic clustering, multimodal fusion, or geographically grounded objectives.

Beyond topic modeling, heterogeneous network embedding methods have been widely studied in domains such as recommendation systems, where multiple node and edge types are jointly modeled within a unified graph representation~\cite{shi2016heterogeneous,zhang2018recommendation}. These methods learn relational embeddings from heterogeneous graphs but are typically optimized for supervised or semi-supervised prediction objectives, rather than unsupervised topic discovery. As a result, they do not explicitly optimize for cluster-level semantic coherence, spatial grounding, or interpretability, which are central to geo-social topic modeling.

Across topic modeling, spatio-semantic extensions, and graph-based representation learning, prior work has largely addressed semantic structure, spatial information, and relational modeling in isolation. Classical and embedding-based topic models focus on semantic similarity, while spatially grounded topic models typically introduce geographic information as auxiliary metadata or enforce it as a post hoc constraint on semantically derived clusters~\cite{Yuan2013Who,Hong2012Discovering}. In parallel, recent graph-based multimodal fusion and contrastive learning methods emphasize learning robust relational embeddings, but are generally not designed for clustering-oriented topic discovery. Consequently, semantic structure, geographic proximity, and clustering objectives are rarely optimized jointly within a single end-to-end framework. Our framework addresses this gap by explicitly modeling semantic and geographic relations as parallel graph structures and integrating them within a unified learning objective tailored for unsupervised topic discovery, yielding clusters that are both semantically coherent and geographically grounded.

\section{Methodology}\label{sec3}

Our approach leverages a collection of geotagged social media posts, where each post (node) $i$ is represented by a semantic embedding $\mathbf{h}_i$, derived from its textual content, and a geographic coordinate $\mathbf{l}_i$. Semantic representations are obtained using Sentence-BERT (SBERT), which fine-tunes transformer-based language models in a Siamese architecture to directly optimize sentence-level similarity. SBERT has been shown to effectively capture semantic proximity in short, informal texts such as tweets \cite{Reimers2019}. To enable joint reasoning over heterogeneous data types, all modalities are embedded into graph-based representations that serve as a common computational substrate. As already mentioned, graphs provide a flexible framework for modeling local neighborhood structure and relational dependencies induced by semantic similarity and geographic proximity. While this study focuses on semantic and geographic signals, the methodology is formulated at the graph level and is therefore modality-agnostic, assuming additional information sources can be easyly encoded as node features or relational graphs. We explore two complementary graph construction strategies: a unified MonoGraph formulation and a decoupled MultiGraph formulation.

\begin{figure}[t!] 
  \centering
    \vspace{0pt}  
    \includegraphics[ height=12cm, keepaspectratio]{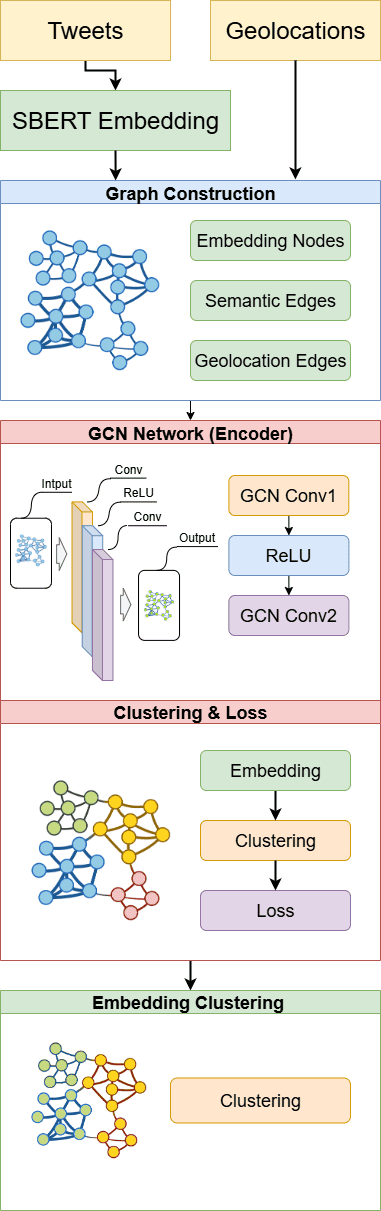}  
    \hspace{1.2cm}  
    \includegraphics[height=12cm, keepaspectratio]{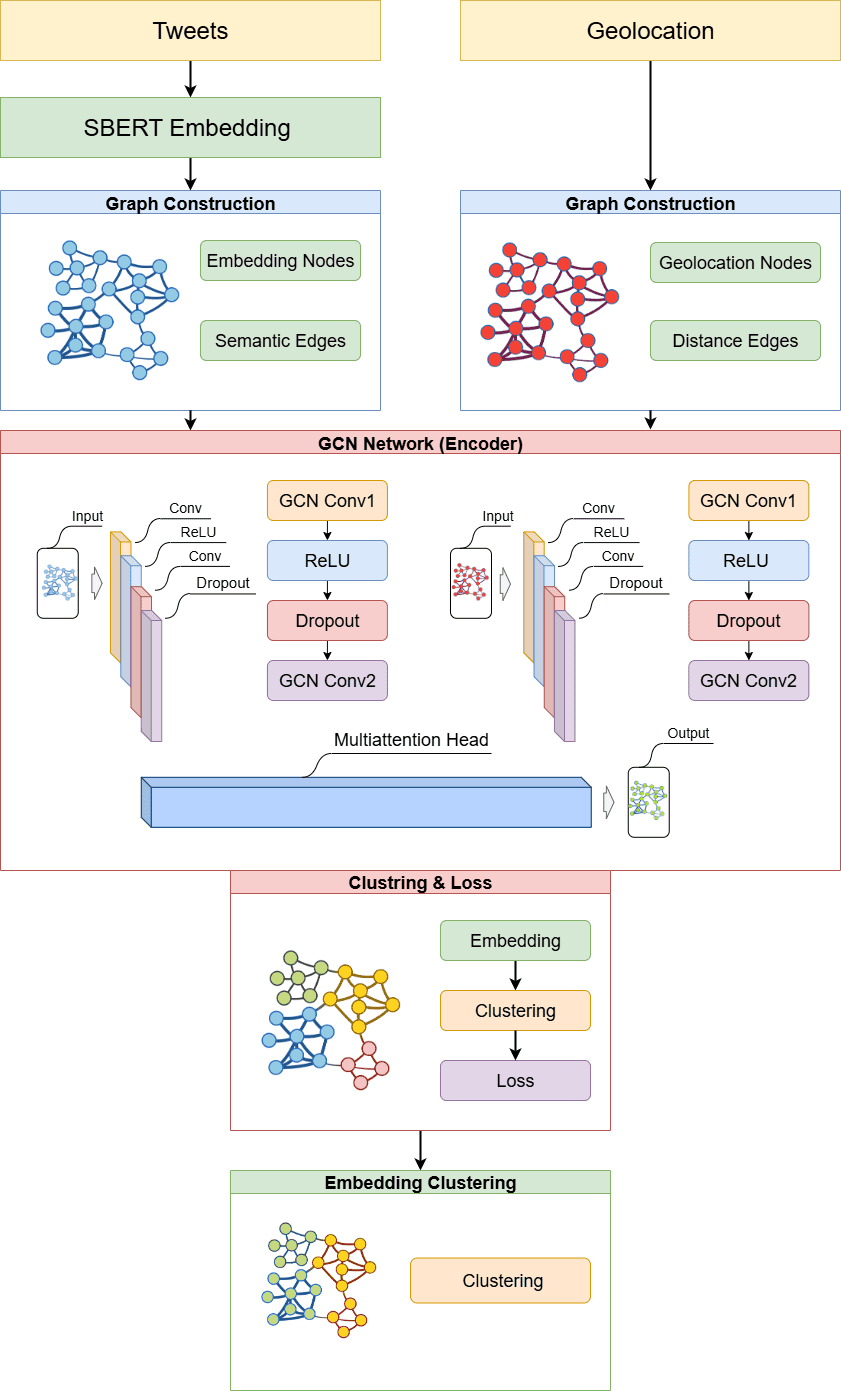}  
    \vspace{0.5cm} 
    \begin{minipage}{0.45\textwidth}
    \vspace{0.3cm} 
    \hspace{-1.8cm}
        \centering
        \textbf{(a)}  
    \end{minipage}
    \hspace{0.7cm}  
    \begin{minipage}{0.45\textwidth}
    \vspace{0.3cm} 
    \hspace{-1.6cm}    
        \centering
        \textbf{(b)}  
    \end{minipage}
    \caption{Layout of the two models: MonoGraph  (a), and MultiGraph (b)}\label{fig:models}
\end{figure}

In the MonoGraph formulation, we construct a single heterogeneous graph $\mathcal{G}_u = (\mathcal{V}, \mathcal{E}_u)$ over social media posts, where each node $i \in \mathcal{V}$ corresponds to a post represented by a semantic embedding $\mathbf{h}_i$ and geographic coordinates $\mathbf{p}_i = (\phi_i, \lambda_i)$.
The graph captures semantic and geographic information through two relation types defined over a shared node set. Semantic similarity between nodes is measured using cosine similarity, defining
a semantic edge type with weights
\[
w_{ij}^{(s)} = \cos(\mathbf{h}_i, \mathbf{h}_j).
\]

Geographic proximity is measured using the Haversine distance
\[
d_{\mathrm{hav}}(\mathbf{p}_i, \mathbf{p}_j) =
2 R \arcsin\!\left(
\sqrt{
\sin^2\!\left(\frac{\phi_j - \phi_i}{2}\right) +
\cos(\phi_i)\cos(\phi_j)
\sin^2\!\left(\frac{\lambda_j - \lambda_i}{2}\right)
}
\right),
\]
where $R \approx 6371\,\mathrm{km}$ denotes the Earth’s radius. Geographic proximity
defines a geographic edge type with weights
\[
w_{ij}^{(g)} = \frac{1}{1 + d_{\mathrm{hav}}(\mathbf{p}_i, \mathbf{p}_j)}.
\]

For each node $i$, a $k$-nearest-neighbor set is determined based on semantic similarity. Both semantic and geographic edge types are instantiated over this shared neighborhood, without introducing explicit modality weighting. Graph sparsity is controlled solely through $k$-nearest neighbor selection. The resulting heterogeneous graph is processed by a heterogeneous GCN encoder composed of relation-specific graph convolution operators. At each layer, messages from different relations are aggregated via summation, enabling joint propagation of semantic similarity and spatial proximity within a unified message-passing framework.

In the MultiGraph formulation, semantic and geographic relations are instantiated as two independent graphs over the same node set $\mathcal{V}$, allowing modality-specific neighborhood structures to be modeled separately. The semantic graph $\mathcal{G}_s = (\mathcal{V}, \mathcal{E}_s)$ connects each node to its
top-$k$ nearest neighbors based on cosine similarity of semantic embeddings, using the semantic edge weights $w_{ij}^{(s)}$ defined above. Graph sparsity is controlled solely through $k$-nearest-neighbor selection, with no global similarity threshold. In parallel, the geographic graph $\mathcal{G}_g = (\mathcal{V}, \mathcal{E}_g)$ is constructed by connecting each node to its $k$ geographically nearest neighbors based on great-circle distance, with geographic edge weights $w_{ij}^{(g)}$ computed using the Haversine distance defined above. No additional thresholds or modality-specific reweighting are introduced. Each graph is processed by a dedicated GCN encoder, enabling semantic and geographic neighborhood information to be learned independently prior to fusion.

\subsection{Network Architectures}\label{3.1}

A schematic overview of both architectures is provided in Fig.~\ref{fig:models}. We employ Graph Convolutional Networks (GCNs) to model local neighborhood interactions induced by semantic and geographic graph structures. GCNs operate via a localized message-passing scheme, updating node features according to
\[
    \mathbf{X}^{(l+1)} = \sigma \left( \hat{A} \mathbf{X}^{(l)} \mathbf{W}^{(l)} \right),
\]
where $\hat{A}$ is the normalized adjacency matrix, $\mathbf{X}^{(l)}$ denotes the feature matrix at layer $l$, $\mathbf{W}^{(l)}$ is a learnable weight matrix, and $\sigma$ is a non-linear activation function such as ReLU. In both architectures, we use two GCN layers with ReLU activations. In the MultiGraph setup, dropout and layer normalization are applied between layers to improve generalization and training stability.

In the MonoGraph architecture, the unified graph is processed by a single shared GCN encoder, yielding joint node embeddings that integrate semantic similarity and geographic proximity through shared message passing. In contrast, the MultiGraph architecture employs two separate GCN branches, one operating on the semantic graph and one on the geographic graph, producing semantic embeddings $\mathbf{x}_i^{(s)}$ and geographic embeddings $\mathbf{x}_i^{(g)}$ for each node $i$. These modality-specific embeddings are subsequently fused through a learnable multi-head cross-attention mechanism. In this formulation, geographic embeddings serve as the queries $(Q)$, while semantic embeddings act as both the keys $(K)$ and values $(V)$. Since attention is applied independently at the node level rather than over sequences, this mechanism functions as a geo-conditioned feature gating operation, allowing the model to selectively reweight semantic feature dimensions based on geographic context rather than modeling interactions across nodes. This asymmetric design is intentional: geographic context provides a stable, low-dimensional conditioning signal that modulates which semantic features are emphasized, rather than being transformed itself. Treating geographic embeddings as queries allows the model to adaptively select and reweight semantic dimensions that are most relevant for a given spatial context, while preserving the expressive capacity of the semantic representations. Formally, the fused representation is obtained as:
\[
\mathbf{z}_i = \mathrm{Attn}\bigl(\mathbf{x}_i^{(g)}, \mathbf{x}_i^{(s)}, \mathbf{x}_i^{(s)}\bigr),
\]
where $\mathrm{Attn}(\cdot)$ denotes multi-head scaled dot-product attention. The multi-head attention layer first linearly projects $(Q, K, V)$ into $h$ subspaces (attention heads) of dimension $d_k = \frac{\text{embed\_dim}}{h}$. For each head, scaled dot-product attention is computed as:
\[
\mathrm{Attention}(\mathbf{Q}, \mathbf{K}, \mathbf{V}) =
\mathrm{softmax}\!\left(\frac{\mathbf{Q}\mathbf{K}^\top}{\sqrt{d_k}}\right)\mathbf{V},
\]
\vspace{2mm}

\noindent and the outputs of all heads are concatenated and linearly transformed to produce the final fused embedding.

\subsection{Loss Design and Training Objective}\label{sec3.2}

To guide the learning of node representations, we optimize a composite objective that combines contrastive, coherence, and alignment losses. The contrastive loss preserves local semantic neighborhood structure induced by the graph, while the coherence and alignment losses introduce global structural regularization by encouraging compact and well-separated clusters in the embedding space. Joint optimization of these objectives enables the model to learn representations that capture both local relational consistency and global cluster organization. During training, clustering is incorporated to provide coarse-grained global guidance that complements the local nature of contrastive learning. At each training epoch, node embeddings are clustered using the K-means algorithm with a fixed number of clusters ($k_{\text{means}} = 15$) and a fixed random seed ($42$) to ensure consistency across epochs. The resulting cluster assignments are used exclusively to compute the coherence and alignment losses and act as dynamic pseudo-labels, while the contrastive loss operates solely on local graph neighborhoods. Since clustering is non-differentiable and performed outside the computational graph, it does not impose hard constraints on the embedding geometry. For evaluation and qualitative analysis, clustering is performed post hoc on the final learned embeddings using Spectral Clustering, as described in Section~\ref{sec3.3}.

\noindent\textbf{Coherence Loss:}
The coherence loss consists of two complementary components: intra-cluster similarity, which promotes internal consistency within clusters, and inter-cluster similarity, which discourages overlap between different clusters.

The intra-cluster similarity measures how similar the embeddings of nodes within the same cluster are to each other:
\[
\mu_{\text{intra}} =
\frac{
\sum_{k=1}^{K}
\sum_{\substack{i,j \in C_k \\ i \neq j}}
\cos(\mathbf{z}_i, \mathbf{z}_j)
}{
\sum_{k=1}^{K} \binom{|C_k|}{2}
},
\]
where \(K\) is the number of clusters and \(C_k\) denotes the set of nodes assigned to cluster \(k\).
Cosine similarity is computed as
\[
\cos(\mathbf{z}_i, \mathbf{z}_j) =
\frac{\mathbf{z}_i \cdot \mathbf{z}_j}
{\|\mathbf{z}_i\|_2 \|\mathbf{z}_j\|_2}.
\]

\vspace{2mm}

\noindent Maximizing intra-cluster similarity encourages compact and internally consistent clusters in the learned embedding space.

The inter-cluster similarity captures the similarity between embeddings belonging to different clusters:
\[
\mu_{\text{inter}} =
\frac{
\sum_{\substack{i,j \in P_{\text{inter}} \\ C(i) \neq C(j)}}
\cos(\mathbf{z}_i, \mathbf{z}_j)
}{
\sum_{i \neq j} \mathbf{1}_{\{C(i) \neq C(j)\}}
},
\]
\vspace{2mm}

\noindent where \(P_{\text{inter}}\) denotes all node pairs drawn from distinct clusters. Lower values indicate better separation between clusters. To jointly encourage intra-cluster compactness and inter-cluster separation, the coherence loss is defined as:

\[ \mathcal{L}_{\text{coherence}} = -\left( \frac{\sum_{k=1}^{K} \sum_{\substack{i,j \in C_k \\ i \neq j}} \cos(\mathbf{z}_i, \mathbf{z}_j)}{\sum_{k=1}^{K} \binom{|C_k|}{2}} - \lambda \cdot \frac{\sum_{\substack{i,j \in P_{\text{inter}} \\ C(i) \neq C(j)}} \cos(\mathbf{z}_i, \mathbf{z}_j)}{\sum_{i \neq j} \mathbf{1}_{\{C(i) \neq C(j)\}}} \right), \]

\noindent where \(\lambda\) controls the trade-off between the two terms.

\vspace{3mm}

\noindent\textbf{Contrastive Loss:}
To preserve local semantic or structural neighborhoods, we employ a contrastive loss that pulls embeddings of similar nodes closer while pushing dissimilar ones apart. Positive pairs are constructed from the semantic graph: for each node \(i\), its positive set consists of its top-$k$ semantic neighbors connected to \(i\) by edges in the semantic graph, while all other nodes within the same mini-batch act as implicit negatives. The contrastive objective is defined as:
\[
\mathcal{L}_{\text{contrast}} =
-\sum_{(i,j) \in \mathcal{P}}
\log
\frac{
\exp(\cos(\mathbf{z}_i, \mathbf{z}_j) / \tau)
}{
\sum_{(i,k) \in \mathcal{N}_i}
\exp(\cos(\mathbf{z}_i, \mathbf{z}_k) / \tau)
},
\]
where \(\mathcal{P}\) denotes the set of positive node pairs, \(\mathcal{N}_i\) contains both positive and negative samples for node \(i\), and \(\tau\) is a temperature parameter controlling the sharpness of the similarity distribution. Unless otherwise stated, we use a fixed temperature parameter $\tau = 0.5$
during training, as identified by the sensitivity analysis in
Section~\ref{sec:ablation_sensitivity}.

\vspace{3mm}

\noindent\textbf{Alignment Loss:}
The alignment loss encourages node embeddings to remain close to the centroid of their assigned cluster:
\[
\mathcal{L}_{\text{align}} =
\frac{1}{N}
\sum_{i=1}^{N}
\left\|
\mathbf{z}_i - \mathbf{c}_{C(i)}
\right\|_2^2,
\]
where \(\mathbf{c}_{C(i)}\) denotes the centroid of cluster \(C(i)\), computed as the mean embedding of all nodes assigned to that cluster.

\vspace{3mm}

\noindent\textbf{Total Loss Objective:}
The final training objective combines all components as:
\[
\mathcal{L}_{\text{total}} =
\alpha \cdot \mathcal{L}_{\text{contrast}} +
\beta \cdot \mathcal{L}_{\text{coherence}} +
\gamma \cdot \mathcal{L}_{\text{align}},
\]
where \(\alpha\), \(\beta\), and \(\gamma\) control the relative contribution of each loss term. Unless otherwise stated, we use the baseline configuration
\(\alpha = 0.8\), \(\beta = 0.2\), \(\gamma = 0.1\), \(k_{\text{means}} = 15\), \(seed = 42\), and \(\tau = 0.5\), during training, as identified by the sensitivity analysis in this section.

\subsection{Embedded Features Clustering}\label{sec3.3}

After obtaining the embedded features produced by the final layer of the GCN encoder, we apply unsupervised clustering to the resulting embeddings \(\mathbf{z}_i\). Clustering helps to reveal latent topic or community structures in the joint semantic-geographic representation. We use Spectral Clustering, which first computes a similarity matrix that reflects the pairwise relationships between data points, and then utilizes the eigenvectors of the graph Laplacian to partition the data. This method is particularly effective for high-dimensional and non-convex data distributions, such as those found in graph-based embeddings \cite{von2007}. We fix the number of clusters ($k_{\text{spect}} = 10$) for post hoc analysis to stabilize the clustering process and reduce variability, ensuring consistent topic identification across runs. This choice is made independently of the training-time clustering used for representation learning and reflects a trade-off between topic granularity and interpretability commonly adopted in topic modeling evaluation. Fixing $k_{\text{spect}}$ also enhances reproducibility by maintaining a consistent clustering configuration.

\section{Experimental Setup}\label{sec4}

This section describes the datasets, preprocessing steps, model configuration, and training protocol used throughout the experimental evaluation. All design choices are reported once here to ensure consistency and reproducibility across subsequent analyses, including clustering evaluation and Topic Quality (TQ) benchmarking.

\subsection{Datasets}\label{sec4.1}

As the models are designed specifically for the analysis of social media posts in disaster situations, we validated our proposed methods using four tweet datasets corresponding to distinct real-world disaster events:  
(1) the 2014 Napa earthquake~\cite{brocher2015}, which occurred on August 24, 2014. The geographic bounds focus on the Napa Valley region in California, covering 4,287 km\(^2\);  
(2) Hurricane Harvey in 2017~\cite{noaa2017}, a Category 4 hurricane impacting southeastern Texas between August 25 and September 7, 2017, covering a spatial extent of 26,736 km\(^2\);  
(3) the 2021 Ahr Valley flood~\cite{koks2021}, which occurred in July 2021 and affected a 1,256 km\(^2\) region in Germany; and  
(4) the 2023 Chile wildfires~\cite{cordero2024}, which impacted central and southern Chile in February 2023 and span a large area of 546,905 km\(^2\). All datasets were collected using Twitter’s former v1.1 streaming and recent search API endpoints, filtered by geographic bounding polygons and event-specific time frames \footnote{The data supporting the findings of this study are available from the authors upon reasonable request, subject to platform terms and conditions.} . Table~\ref{tab:dataset} summarizes the key properties of each dataset.

\begin{table}[t]
\centering
\caption{Properties of the datasets used for validation.}
\label{tab:dataset}
\setlength{\tabcolsep}{2.8pt}
\renewcommand{\arraystretch}{1.5}
\begin{tabular}{@{}lcccc@{}}
\toprule
\textbf{Dataset} & \textbf{Posts} & \textbf{Time Frame} & \textbf{Geographic Bounds} & \textbf{Cover Area} \\ \midrule
Ahr Valley floods & 11,177 & 2021-07-01 – 2021-07-31 & (50.01, 6.00, 50.75, 8.00) & 1,256 km$^2$ \\
Hurricane Harvey  & 28,933 & 2017-08-25 – 2017-09-07 & (27.58, –98.77, 30.66, –90.86) & 26,736 km$^2$ \\
Chile wildfires   & 100,707 & 2023-02-01 – 2023-02-28 & (–55.95, –76.89, –17.34, –65.22) & 546,905 km$^2$ \\
Napa earthquake   & 90,907 & 2014-08-24 – 2014-08-25 & (37.11, –123.05, 38.98, –121.04) & 4,287 km$^2$ \\ 
\bottomrule
\end{tabular}
\end{table}

\subsection{Preprocessing}\label{sec4.2}

As is standard practice in natural language processing, the raw tweet text was preprocessed prior to analysis to ensure consistency and reduce noise. URLs were removed using regular expressions and all text was lowercased. HTML entities were unescaped, user mentions were removed, and excess whitespace was stripped. Non-alphanumeric characters and numeric tokens were eliminated. For embedding generation, we follow the input preprocessing conventions recommended for Sentence-BERT, replacing mentions with the token \texttt{@user}, mapping URLs to \texttt{http}, and removing newline characters. These steps ensure compatibility with the pre-trained language models and consistency across datasets.

\subsection{Model Configuration and Training}\label{sec4.3}

All models were trained in an unsupervised manner without access to ground-truth topic or cluster labels, using the compound objective described in Section~\ref{sec3.2}. Input features consisted of SBERT-generated semantic embeddings and associated geolocation coordinates, which were used to construct semantic and geographic graphs in both the MonoGraph and MultiGraph formulations. Training was performed using the Adam optimizer with an initial learning rate of \(1\times10^{-3}\). Models were trained for 50 epochs with early stopping based on coherence stability on a held-out validation subset. In the MultiGraph configuration, each branch employed two GCN layers with a dropout rate of 0.3. Fusion was performed via multi-head attention with four heads, and layer normalization was applied between fusion stages to stabilize training.

\subsection{Scalability via Strided Chunking}\label{sec4.4}

To scale graph construction to large corpora, we adopt strided chunking during preprocessing. Given a stride \(s\), the dataset is partitioned into \(s\) partially overlapping subsets defined as
\[
\text{chunk}_i = \text{Input}_{\mathrm{gdf}}[i::s], \quad i = 1, \dots, s.
\]
This construction ensures that each node appears in exactly one position per stride cycle while nodes near chunk boundaries are shared across multiple chunks, preserving local semantic and geographic neighborhoods. Chunk overlap prevents artificial fragmentation of neighborhoods that could otherwise arise from independent graph construction. As a result, node embeddings remain consistent across chunks, and striding affects computational efficiency only, not the learned representation or clustering behavior. Full data coverage was verified prior to training, and all chunks were processed sequentially under identical optimization settings. Empirically, varying the stride within reasonable bounds did not result in meaningful
changes in the embedding geometry or TQ, confirming that strided chunking functions strictly as a scalability mechanism and does not bias embedding geometry or topic structure. Stride values are selected based on memory constraints rather than dataset-specific tuning, and do not affect neighborhood structure or learned representations.

\subsection{Baseline Configuration}\label{sec4.5}

The primary baseline for all experiments is the sequential semantic workflow introduced in this study. It operates exclusively on semantic embeddings and does not incorporate graph-based modeling, geographic information, multimodal fusion, or contrastive learning. This configuration serves as a minimal reference point and enables direct attribution of performance differences to the proposed graph construction, semantic--geographic integration, and contrastive objectives. In addition, we include the JSTTS multimodal pipeline proposed by Hanny et al.~\cite{hanny2025} as an external reference method. Although temporal and sentiment features are part of the original JSTTS formulation, these modalities are intentionally excluded from all primary experiments in this work. This design choice applies uniformly across all evaluated models and is motivated by the goal of isolating the methodological contribution of semantic--geographic graph modeling and contrastive learning. Incorporating auxiliary temporal or affective signals would alter the input space and introduce additional information that is orthogonal to the methodological focus of this study, thereby confounding attribution of performance differences. Nonetheless, for completeness and transparency, the behavior of JSTTS under its original full-modality configuration is reported separately in the evaluation section. These results are provided for contextual reference only and are not used for direct comparison with the proposed models.

To ensure comparability with prior work, we use the same datasets and embedding backbones as in~\cite{hanny2025} across all experiments, including clustering and topic quality evaluation. Specifically, the English \texttt{all-MiniLM-L12-v2} model is used for the Napa earthquake and Hurricane Harvey datasets, while the multilingual \texttt{distiluse-base-multilingual-cased-v1} model is used for the Ahr Valley floods and Chile wildfires datasets. All models are evaluated using identical preprocessing pipelines, clustering procedures, and keyword extraction methods.


\section{Evaluation}\label{sec5}

This section evaluates the proposed models from two complementary perspectives. First, we assess their ability to produce semantically coherent and well-separated clusters in the embedding space. Second, we benchmark topic modeling performance using TQ against a sequential baseline and an established multimodal pipeline. Together, these evaluations quantify both representation quality and downstream topic modeling effectiveness across diverse disaster scenarios.

\begin{table}[!t]
\caption{Clustering evaluation results across different datasets using the two models against the input referenced data}\label{tab1}
\label{tab:clustering}
\setlength{\tabcolsep}{7pt} 
\renewcommand{\arraystretch}{1.5}
\begin{tabular}{@{}lcccccccc@{}}
\toprule
Dataset & \multicolumn{2}{c}{Ahr Valley Floods} & \multicolumn{2}{c}{Hurricane Harvey} & \multicolumn{2}{c}{Chile Wildfires} & \multicolumn{2}{c}{Napa Earthquake} \\ \midrule
Model   & Inter     & Intra     & Inter     & Intra    & Inter    & Intra    & Inter    & Intra    \\\midrule
Input        & 0.432          & 0.779          & 0.401          & 0.693          & 0.355          & 0.525          & 0.319          & 0.447          \\
MonoGraph  & \textbf{0.046} & 0.846          & \textbf{0.045} & \textbf{0.817} & -0.071          & 0.777          & -0.054          & 0.774          \\
MultiGraph  & 0.248          & \textbf{0.853} & 0.164          & 0.824          & \textbf{-0.002} & \textbf{0.790} & \textbf{-0.042} & \textbf{0.791} \\ \bottomrule
\end{tabular}
\end{table}

\begin{figure}[t!]
    \vspace{-2mm}
    \centering
    \begin{subfigure}[t]{0.4\textwidth}
        \includegraphics[width=\linewidth]{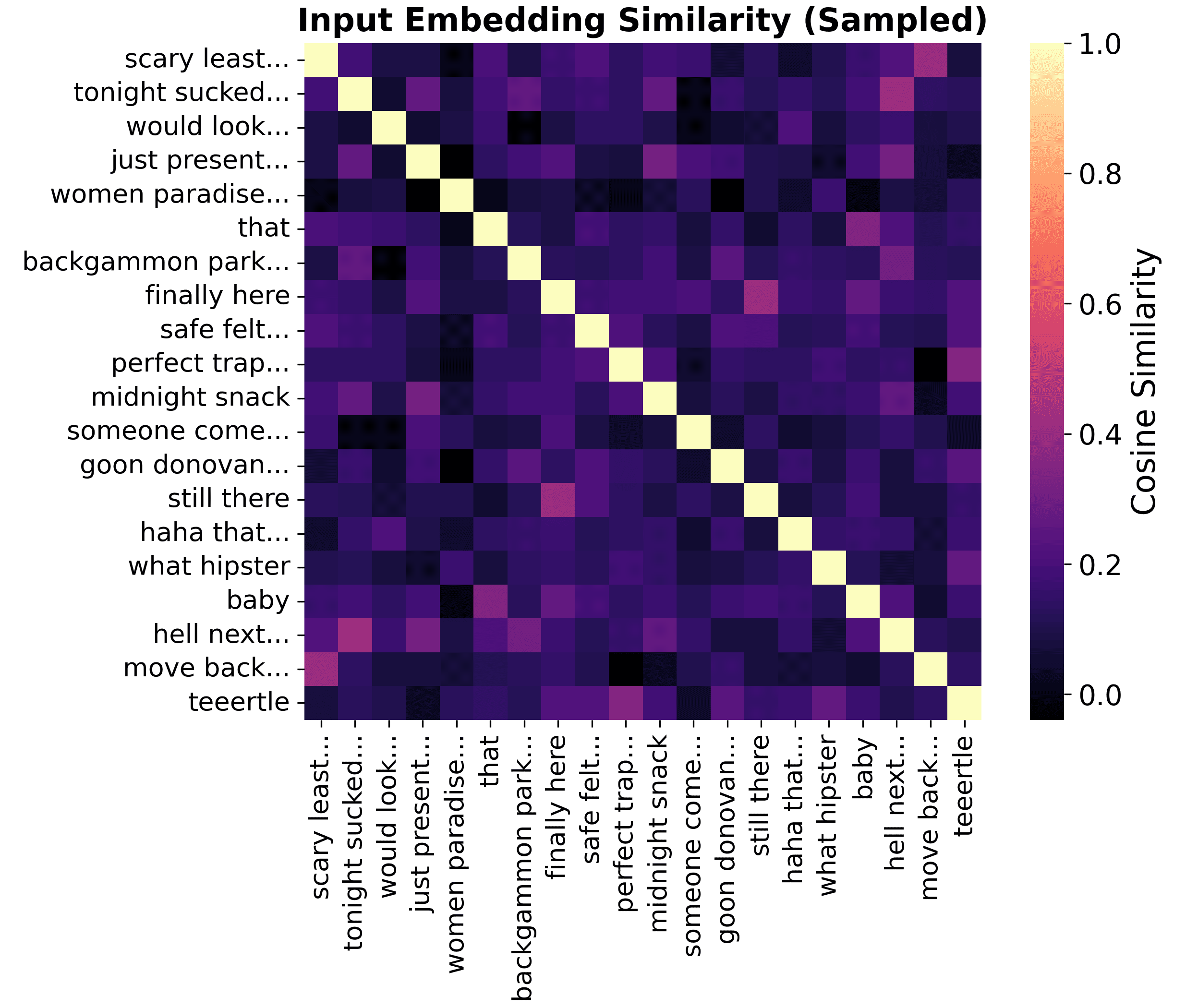}
        \caption{}
    \end{subfigure}
    \begin{subfigure}[t]{0.4\textwidth}
     \includegraphics[width=\linewidth]{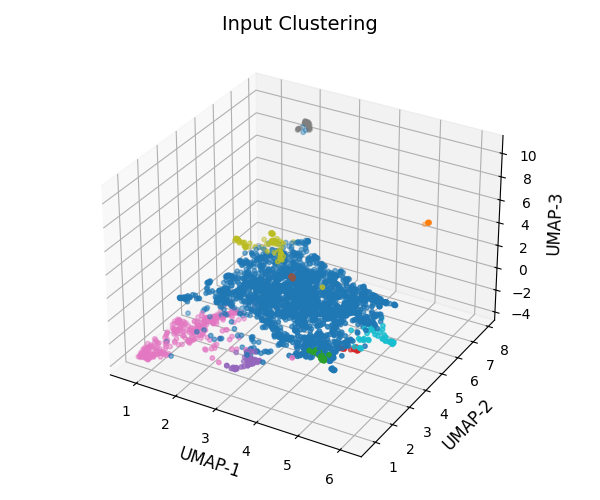}
        \caption{}
    \end{subfigure}
    \hfill     
\\
    \begin{subfigure}[t]{0.4\textwidth}
        \includegraphics[width=\linewidth]{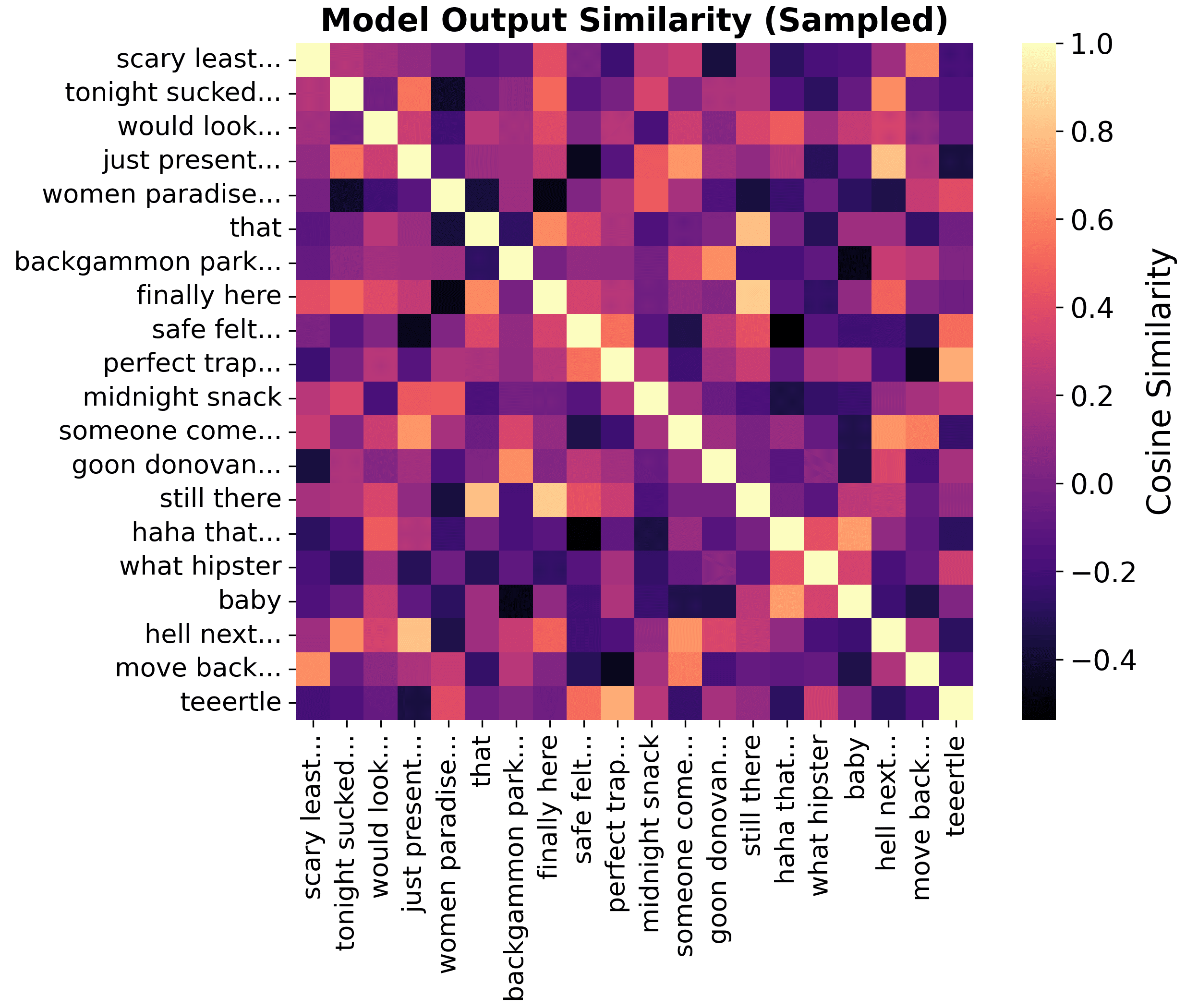}
        \caption{}
    \end{subfigure}
    \begin{subfigure}[t]{0.4\textwidth}
        \includegraphics[width=\linewidth]{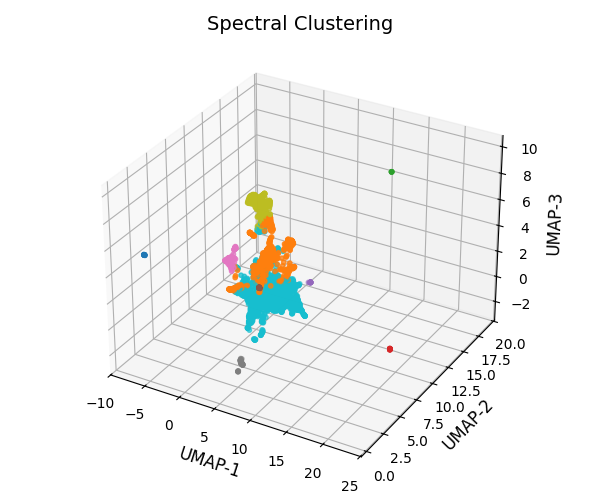}
        \caption{}
    \end{subfigure}
    \hfill
\\    
    \begin{subfigure}[t]{0.4\textwidth}
        \includegraphics[width=\linewidth]{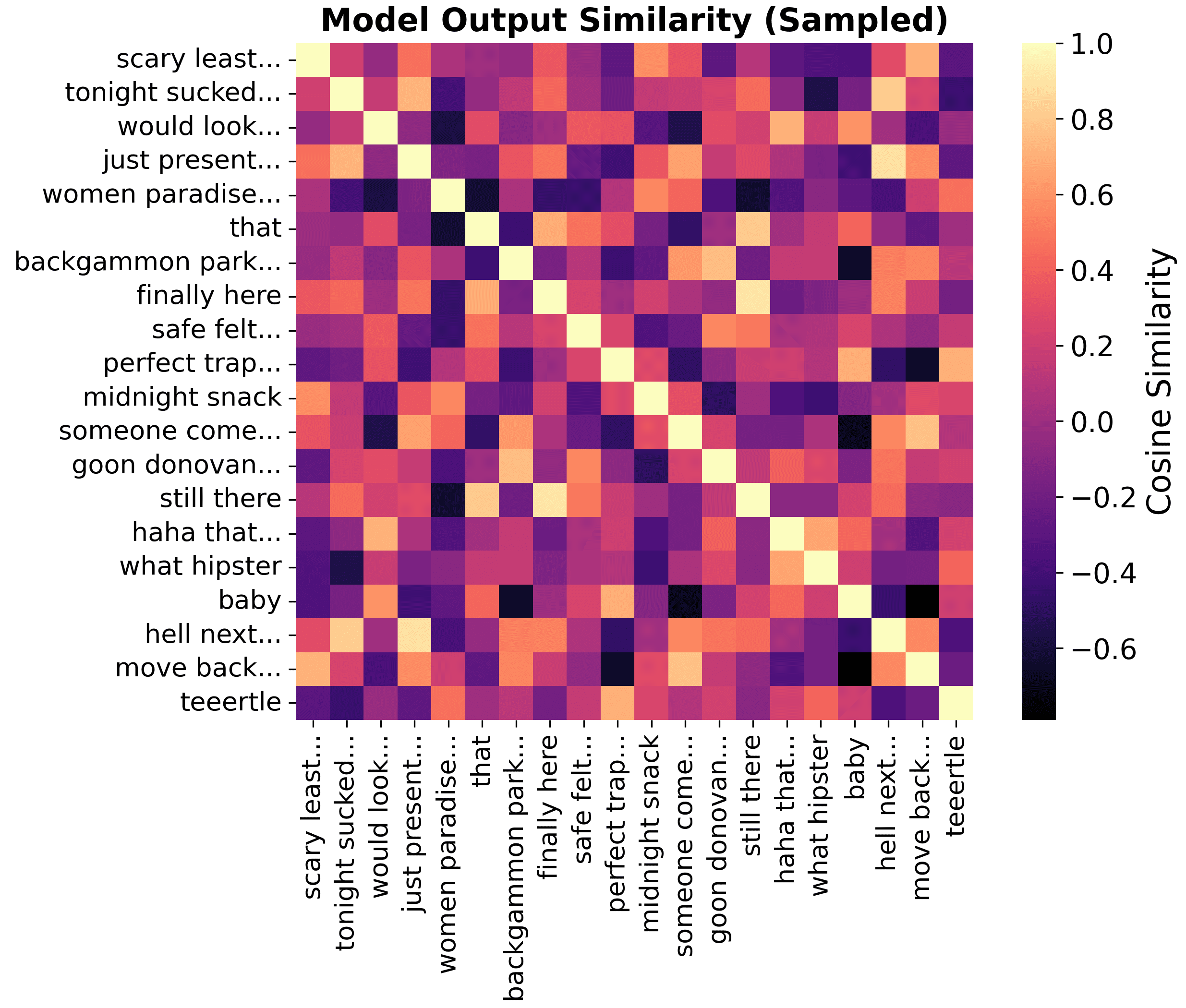}
        \caption{}
    \end{subfigure}
    \begin{subfigure}[t]{0.4\textwidth}
        \includegraphics[width=\linewidth]{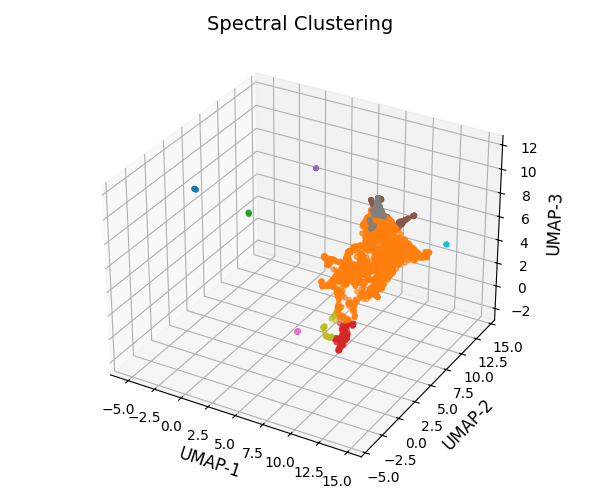}
        \caption{}
    \end{subfigure}
    \caption {Similarity matrices and their corresponding 3D clustering visualizations (after dimensionality reduction) for a subset of the Napa dataset. The first row (a, b) shows the input embeddings, while the second row (c, d) and third row (e, f) display the results for MonoGraph and MultiGraph models, respectively.}

    \label{fig:clustering} 
    \vspace{-2mm}
\end{figure}

\subsection{Clustering Capability Evaluation}\label{sec5.1}

To evaluate the models' capacity for generating semantically coherent representations, we conducted clustering assessments on both the original input embeddings and the output embedded features produced by each model. Following the methodology outlined in Section~\ref{sec3.2}, we computed two complementary metrics: intra-cluster similarity, reflecting semantic cohesion within clusters, and inter-cluster similarity, indicating the degree of separation between clusters. Effective clustering is characterized by high intra-cluster cohesion and low inter-cluster similarity.

Table~\ref{tab:clustering} summarizes the results across four disaster-specific datasets. The raw input embeddings serve as a reference baseline and consistently exhibit moderate intra-cluster cohesion but relatively poor inter-cluster separation, indicating that semantic structure alone is insufficient for robust clustering. Both MonoGraph and MultiGraph substantially improve clustering behavior across all datasets. Inter-cluster similarity decreases markedly, while intra-cluster cohesion increases, indicating tighter and better-separated clusters. MultiGraph achieves the highest intra-cluster cohesion across all datasets and yields the lowest (i.e., most negative) inter-cluster similarity for the Chile wildfires and Napa earthquake datasets, suggesting particularly well-isolated and internally consistent clusters. For the Ahr Valley floods and Hurricane Harvey datasets, MonoGraph attains slightly lower inter-cluster similarity values, although its intra-cluster cohesion remains marginally below that of MultiGraph while still substantially outperforming the baseline.

These quantitative findings are reinforced by the visualizations in Fig.~\ref{fig:clustering}. The input similarity matrix exhibits diffuse structure, whereas the learned representations display sharper contrast and clearer block patterns. Likewise, the 3D visualizations reveal a transition from scattered point clouds to compact and well- separated clusters. Overall, both models, and especially MultiGraph, demonstrate a strong ability to reorganize the embedding space in a way that enhances clustering quality, which is critical for downstream topic modeling and spatial-semantic analysis.

\subsection{Topic Quality Benchmarking}\label{sec5.2}

This section evaluates the semantic quality of the topics discovered by the proposed models under controlled and comparable conditions. We first describe the evaluation protocol and baseline configurations used for TQ assessment, then formally define the TQ metric, and finally present quantitative results across all datasets.

\subsubsection{Evaluation Protocol}

Following the baseline definitions established in Section~\ref{sec4.5}, TQ evaluation is conducted under controlled and directly comparable conditions. All models are evaluated using semantic embeddings as the common input modality. Both the multimodal models additionally incorporate geographic information, while the BERTopic baseline operates solely on semantic representations, reflecting its original formulation. In addition to the primary benchmarking experiments, we report TQ scores for the JSTTS pipeline proposed by Hanny et al.~\cite{hanny2025} in its original intended configuration, with all available modalities, including semantic, spatial, temporal, and sentiment features, enabled (reported in Table~\ref{tab:jstts_tq}). As already mentioned in Section~\ref{sec4.5}, introducing auxiliary information that is orthogonal to the methodological focus of this study. As a result, these results are not treated as directly comparable benchmarks, but are reported for transparency and contextual reference, illustrating the relative scale of TQ achieved by modular multimodal pipelines.

\subsubsection{Topic Quality Metric}

Topic quality is a widely adopted metric for evaluating topic models, as it jointly captures semantic coherence and lexical diversity. It is defined as the product of Topic Coherence (TC) and Topic Diversity (TD), computed using the top-$k$ keywords per topic (with $k_{\text{means}} = 15$ in all experiments). To extract representative keywords, documents assigned to each cluster are aggregated and a TF--IDF vectorizer is applied locally. The top-$k$ terms with the highest TF--IDF scores are selected as topic descriptors. Topic Coherence is computed using normalized Positive Pointwise Mutual Information (NPPMI)~\cite{bouma2009,jurafsky2023}. Given two words $w_i$ and $w_j$, NPPMI is defined as:
\[
\text{NPPMI}(w_i, w_j) = 
\frac{\max\left(0, \log \frac{P(w_i, w_j)}{P(w_i) P(w_j)}\right)}
{-\log P(w_i, w_j)},
\]
where probabilities are estimated from word co-occurrence counts over the full corpus. The coherence of a topic is computed as the average NPPMI over all $\binom{k}{2}$ keyword pairs, and the overall Topic Coherence is obtained by averaging across all valid topics.

Topic Diversity measures how distinct the discovered topics are by quantifying the proportion of unique keywords across all topics:
\[
\text{TD} = \frac{U}{k \times |T|},
\]
where $U$ denotes the number of unique words among the top-$k$ keywords across all topics, and $|T|$ is the total number of topics. This formulation penalizes repeated keywords and rewards lexical diversity. Finally, TQ is defined as:
\[
\text{TQ} = \text{TC} \times \text{TD},
\]
favoring models that produce topics that are both semantically coherent and lexically diverse. All components of TQ are computed using the same preprocessing and keyword extraction pipeline for all evaluated methods.

\begin{table}[t]
\centering
\caption{Topic Quality(TQ) achieved by the JSTTS pipeline when all modalities (semantic, spatial, temporal, and sentiment) are enabled, as originally specified.}
\label{tab:jstts_tq}
\setlength{\tabcolsep}{64pt}
\renewcommand{\arraystretch}{1.3}
\begin{tabular}{lc}
\toprule
\textbf{Dataset} & \textbf{JSTTS TQ} \\
\midrule
Ahr Valley Floods & 0.165 \\
Hurricane Harvey & 0.191 \\
Chile Wildfires & 0.142 \\
Napa Earthquake & 0.081 \\
\bottomrule
\end{tabular}
\end{table}

\subsection{TQ Results and Analysis}\label{sec4.4}

The TQ scores reported in Table~\ref{tab:tq} clearly demonstrate the effectiveness of our proposed models in generating high-quality topic clusters across diverse disaster-related datasets. The MultiGraph consistently achieves the highest TQ scores in three out of four scenarios, the Ahr Valley floods (0.415), Hurricane Harvey (0.384), and the Napa earthquake (0.195), demonstrating its robustness and adaptability to datasets with varying linguistic, geographic, and topical characteristics. This superior performance can be attributed to its cross-attention mechanism, which enables selective regulation of individual modalities and more effective capture of both global semantic patterns and local topical signals. On the Chile wildfires dataset, however, the MonoGraph slightly outperforms the MultiGraph (0.169 vs. 0.161), with both models performing comparably well overall. This suggests that while the MultiGraph generally excels, the benefits of increased model complexity may vary depending on the nature and quality of the data. In particular, contexts characterized by fewer high-quality labeled signals or more multilingual noise may favor the simpler and more direct MonoGraph architecture, which consistently outperforms baseline methods across all datasets and yields stable, focused clusters despite its relative simplicity.

\begin{table}[]
\caption{TQ for outputs from our two models (MonoGraph and MultiGraph), the method by Hanny et al.~\cite{hanny2025}, and the baseline sequential workflow across four datasets.}
\label{tab:tq}
\setlength{\tabcolsep}{7.5pt} 
\renewcommand{\arraystretch}{1.5}
\begin{tabular}{@{}lclcclc@{}}
\toprule
Dataset & \multicolumn{2}{c}{MonoGraph } & MultiGraph  & \multicolumn{2}{c}{Hanny et al.~\cite{hanny2025}} & Sequential Workfollow \\ \midrule
Ahr Valley Floods & \multicolumn{2}{c}{0.383} & \textbf{0.415} & \multicolumn{2}{c}{0.235} & 0.029 \\
Hurricane Harvey  & \multicolumn{2}{c}{0.368} & \textbf{0.384} & \multicolumn{2}{c}{0.222} & 0.042 \\
Chile Wildfires   & \multicolumn{2}{c}{\textbf{0.169}} & 0.161 & \multicolumn{2}{c}{0.117} & 0.034 \\
Napa Earthquake   & \multicolumn{2}{c}{0.191} & \textbf{0.195} & \multicolumn{2}{c}{0.143} & 0.030 \\ \bottomrule
\end{tabular}
\end{table}

The multimodal pipeline \cite{hanny2025} shows limited performance (TQ ranging from 0.117 to 0.235), likely due to the removal of sentiment and temporal features during evaluation. These components, while integral to the original design, appear essential to the pipeline’s effectiveness. While the model presented in this paper was originally designed to incorporate sentiment and temporal features, these were deliberately neutralized in our evaluation to isolate textual topic modeling performance. The resulting drop in quality highlights the fact that the pipeline’s heavy reliance on auxiliary modalities may limit its effectiveness in settings where such features are unavailable, noisy, or intentionally excluded. The sequential workflow performs the worst overall, with TQ scores falling below 0.05 across all datasets. This approach suffers from treating the stages of the modeling pipeline, embedding, clustering, and keyword extraction, as independent components, which results in fragmented topic structure and poor alignment between cluster semantics and extracted keywords. The absence of feedback or interaction between stages appears to hinder its ability to produce globally meaningful topic representations.

\section{Ablation, Sensitivity, and Stability Analysis}\label{sec:ablation_sensitivity}

To provide a systematic understanding of the proposed multimodal graph-based topic modeling framework, we conduct a comprehensive set of ablation, sensitivity, and stability experiments. These analyses are designed to: (i) isolate the contribution of individual loss components and architectural choices, and (ii) assess the robustness of the model with respect to hyperparameter settings and stochastic initialization. All experiments are conducted on the Hurricane Harvey dataset using both the MonoGraph and MultiGraph variants. Performance is evaluated using Topic Quality, which jointly reflects topic coherence and diversity. Unless otherwise stated, results reported in this section are averaged over the corresponding experimental settings, and summary statistics are provided in the form of Mean, standard deviation (STD), and 95\% Confidence Interval (CI) where applicable.

\begin{table}[t]
\centering
\caption{Loss ablation results based on TQ.
\checkmark\ indicates the loss is included, while $\times$ indicates it is excluded.}
\label{tab:ablation}
\setlength{\tabcolsep}{17pt}
\renewcommand{\arraystretch}{1.4}
\begin{tabular}{lccccc}
\toprule
\textbf{Model} &
$\mathcal{L}_c$ &
$\mathcal{L}_h$ &
$\mathcal{L}_a$ &
\textbf{Removed Loss} &
\textbf{TQ} \\
\midrule
MonoGraph  & \checkmark & \checkmark & \checkmark & None (Full) & \textbf{0.368} \\
MonoGraph  & $\times$   & \checkmark & \checkmark & Contrastive & 0.205 \\
MonoGraph  & \checkmark & $\times$   & \checkmark & Coherence   & 0.357 \\
MonoGraph  & \checkmark & \checkmark & $\times$   & Alignment   & 0.353 \\
\midrule
MultiGraph & \checkmark & \checkmark & \checkmark & None (Full) & \textbf{0.384} \\
MultiGraph & $\times$   & \checkmark & \checkmark & Contrastive & 0.218 \\
MultiGraph & \checkmark & $\times$   & \checkmark & Coherence   & 0.362 \\
MultiGraph & \checkmark & \checkmark & $\times$   & Alignment   & 0.366 \\
\bottomrule
\end{tabular}
\end{table}

\subsection{Loss Ablation Analysis}

We analyze the contribution of each loss component, contrastive loss ($\mathcal{L}_c$), coherence loss ($\mathcal{L}_h$), and alignment loss ($\mathcal{L}_a$), by removing one term at a time while keeping all other hyperparameters fixed at their baseline configuration. Results are summarized in Table~\ref{tab:ablation}. Removing the contrastive loss leads to a substantial degradation in performance for both model variants, with TQ dropping sharply relative to the full configuration. This indicates that contrastive learning provides the primary discriminative signal necessary for effective topic separation. In contrast, removing either the coherence or alignment loss results in only moderate reductions in TQ, suggesting that these components act as stabilizing regularizers rather than dominant drivers of clustering quality. Overall, the ablation results confirm that the loss formulation is well balanced: contrastive learning establishes topic structure, while coherence and alignment losses refine and stabilize the learned representations.

\begin{table}[!t]
\centering
\caption{Sensitivity analysis of loss weights based on Topic TQ.}
\label{tab:sensitivity-loss}
\setlength{\tabcolsep}{42pt}
\renewcommand{\arraystretch}{1.3}
\begin{tabular}{lcc}
\toprule
Setting & MonoGraph & MultiGraph \\
\midrule
$\alpha=0.2$ & 0.308 & 0.346 \\
$\alpha=0.5$ & 0.351 & 0.355 \\
$\alpha=0.8$ & \textbf{0.368} & \textbf{0.384} \\
\midrule
$\beta=0.1$  & 0.344 & 0.363 \\
$\beta=0.2$  & \textbf{0.368} & \textbf{0.384} \\
$\beta=0.5$  & 0.365 & 0.354 \\
\midrule
$\lambda=0.05$ & 0.361 & 0.367 \\
$\lambda=0.1$  & \textbf{0.368} & \textbf{0.384} \\
$\lambda=0.2$  & 0.349 & 0.362 \\
\bottomrule
\textbf{Mean} & 0.354 & 0.367 \\
\textbf{STD}  & 0.019 & 0.014 \\
\textbf{95\% CI} & $\pm$0.013 & $\pm$0.009 \\
\bottomrule
\end{tabular}
\end{table}

\begin{table}[!b]
\centering
\caption{Sensitivity analysis for contrastive temperature $\tau$ and number of clusters ($k_{\text{means}} = 15$) using TQ.}
\label{tab:sensitivity_tau_k}
\setlength{\tabcolsep}{24pt}
\renewcommand{\arraystretch}{1.4}
\begin{tabular}{llcc}
\toprule
Parameter & Value & MonoGraph TQ & MultiGraph TQ \\
\midrule
$\tau$ & 0.2 & 0.314 & 0.359 \\
$\tau$ & 0.5 & \textbf{0.368} & \textbf{0.384} \\
$\tau$ & 1.0 & 0.315 & 0.359 \\
\bottomrule
$k_{\text{means}} = 15$ & 10 & 0.341 & 0.351 \\
$k_{\text{means}} = 15$ & 15 & \textbf{0.368} & \textbf{0.384} \\
$k_{\text{means}} = 15$ & 20 & 0.364 & 0.370 \\
\bottomrule
\textbf{Mean} & -- & 0.345 & 0.368 \\
\textbf{STD}  & -- & 0.026 & 0.017 \\
\textbf{95\% CI} & -- & $\pm$0.021 & $\pm$0.014 \\
\bottomrule
\end{tabular}
\end{table}

\begin{table}[!t]
\centering
\caption{Stability across random seeds computed over independent runs.}
\label{tab:seed}
\setlength{\tabcolsep}{20pt}
\renewcommand{\arraystretch}{1.4}
\begin{tabular}{lcccc}
\toprule
Model & Mean TQ & STD & 95\% CI & Seeds \\
\midrule
MonoGraph  & 0.3667 & 0.0025 & $\pm$0.0062 & \{42,1,2,3\} \\
MultiGraph & 0.3625 & 0.0034 & $\pm$0.0084 & \{42,1,2,3\} \\
\bottomrule
\end{tabular}
\end{table}

\subsection{Hyperparameter Sensitivity Analysis}

We assess robustness to hyperparameter selection by independently varying key parameters while keeping all others fixed at their baseline configuration. Table~\ref{tab:sensitivity-loss} summarizes TQ under different loss-weight configurations. For MonoGraph, the mean TQ across all tested settings is $0.354$ with a standard deviation of $0.019$, while MultiGraph achieves a mean of $0.367$ with a standard deviation of $0.014$. The relatively small dispersion indicates that performance is stable under moderate variations of loss weights, with the baseline configuration consistently yielding the highest TQ. Sensitivity with respect to the contrastive temperature $\tau$ and the number of clusters ($k_{\text{means}} = 15$) is reported in Table~\ref{tab:sensitivity_tau_k}. TQ exhibits a unimodal trend with respect to $\tau$, where intermediate values provide optimal separation of positive and negative pairs. Across all tested values, the standard deviation remains below $0.03$ for both models, indicating limited sensitivity. Similarly, performance remains stable across different values of ($k_{\text{means}}$), with peak TQ achieved at the baseline configuration $k_{\text{means}} = 15$. It is also worth noting that while the absolute topic granularity varies with the choice of $k_{\text{means}}$, the relative performance ranking between MonoGraph and MultiGraph remains stable across the tested range. These results demonstrate that the proposed framework is robust to reasonable variations in both optimization and clustering hyperparameters, and does not rely on finely tuned parameter choices.

\begin{table}[!b]
\centering
\caption{Comparison of fusion strategies for multimodal integration based on TQ.}
\label{tab:fusion}
\setlength{\tabcolsep}{37pt}
\renewcommand{\arraystretch}{1.4}
\begin{tabular}{lcc}
\toprule
Model & Fusion Strategy & TQ \\
\midrule
MultiGraph & Concatenation & 0.225 \\
MultiGraph & Attention  & \textbf{0.384} \\
MultiGraph & Concatenation + MLP  & 0.165 \\
\bottomrule
\end{tabular}
\end{table}

\subsection{Random Seeds Stability and Fusion Strategy}

To evaluate robustness to stochastic initialization, we repeat the full baseline configuration using multiple random seeds. Table~\ref{tab:seed} reports the mean TQ, standard deviation, and 95\% confidence interval. Both models exhibit low variance across runs, indicating that performance is stable and reproducible. We further compare attention-based fusion with alternative fusion strategies for the MultiGraph model, including plain concatenation and concatenation augmented with a multi-layer perceptron (MLP). As shown in Table~\ref{tab:fusion}, attention-based fusion yields substantially higher TQ than both concatenation-based variants, indicating that adaptive, context-aware weighting is more effective for integrating heterogeneous modalities than static or prior-driven fusion schemes.

\subsection{Architectural and graph construction choices}

In addition to the reported ablation and sensitivity analyses, we conducted preliminary validation experiments to assess the impact of architectural depth and graph construction choices on model behavior. In particular, we varied the number of GCN layers and attention heads within a small range (1--3) and observed no consistent or systematic changes in TQ across datasets. Based on this observation, these architectural parameters are fixed in all reported experiments in order to reduce the hyperparameter search space and improve reproducibility. We also explored alternative graph construction strategies, including global similarity and distance-based thresholding. These approaches were found to be sensitive to dataset scale and density, often requiring dataset-specific tuning to maintain stable sparsity levels. To avoid this dependency, we adopt a consistent top-$k$ neighborhood strategy for both semantic and geographic graphs throughout the paper. This choice yields predictable graph sparsity across datasets and allows performance differences to more directly reflect the proposed multimodal fusion mechanism and learning objectives rather than incidental graph construction effects.

\subsection{Overall Model Behavior}

Overall, the ablation, sensitivity, and stability analyses demonstrate that the proposed framework is robust across a wide range of configurations. Performance consistently peaks at the baseline configuration ($\alpha=0.8$, $\beta=0.2$, $\lambda=0.1$,  \(seed = 42\), $k_{\text{means}} = 15$, and $\tau =0.5$) while maintaining low variance across random seeds. These results indicate that the reported improvements are not artifacts of hyperparameter tuning or stochastic effects, but reflect stable and reproducible behavior of the proposed model.

\section{Case Study Evaluation}\label{sec5}

A key motivation behind the design of our models was to ensure interpretability for real-world applications. To complement the experiments in Section \ref{sec4}, we conducted a case study to showcase their utility in a disaster management context. We used a subset of Hurricane Harvey dataset, a tropical storm that occurred in 2017 and hit the state of Texas in the USA on August 25 \cite{noaa2017}. The dataset comprises 28,933 geo-located tweets posted within and around the area of Texas, obtained by filtering tweets based on Geographic Bounds defined by the polygon \(\text{Polygon}(-98.77, 27.58; -90.86, 27.58; -90.86, 30.66; -98.77, 30.66; -98.77, 27.58)\) and the Time Frame from August 25 to September 7, 2017.

\begin{figure}[t!]
    \centering
    \hspace{0.3cm}
    \includegraphics[ width=13cm]{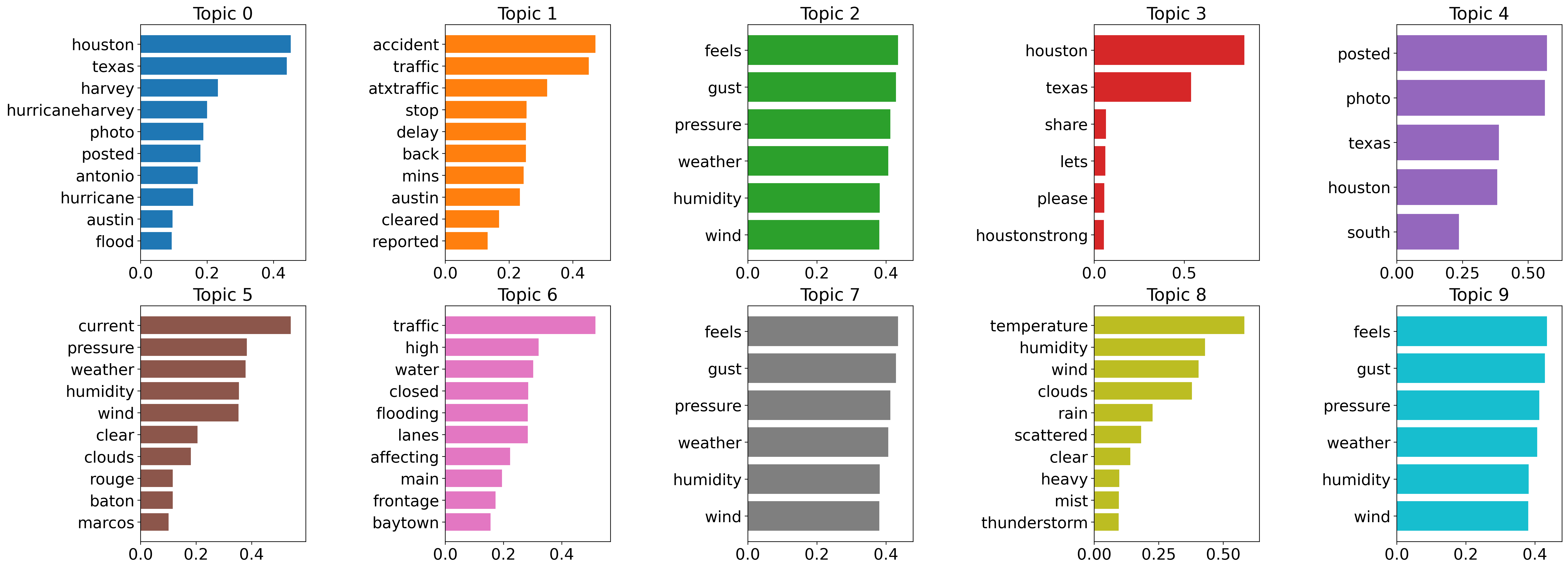}
    \vspace{0.2cm}
    
    \includegraphics[ width=13cm]{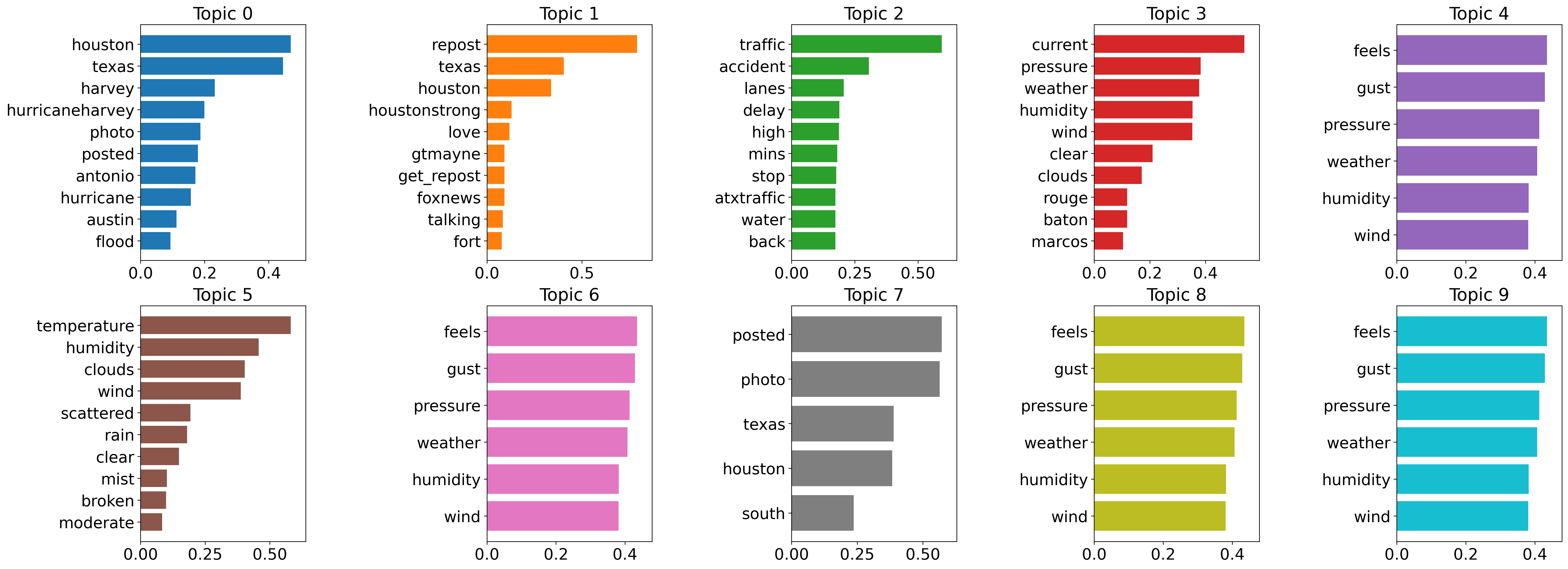}
    
    \vspace{0.4cm} 
    
    \caption{Topic-word distributions for a subset of the Hurricane Harvey dataset, applying MonoGraph  (top) and the MultiGraph  (bottom) models.}\label{fig:keywords}
\end{figure}

\subsection{Topic Interpretation}\label{sec5.1}

\begin{figure}[t!]
    \centering
    \includegraphics[ width=11cm, frame]{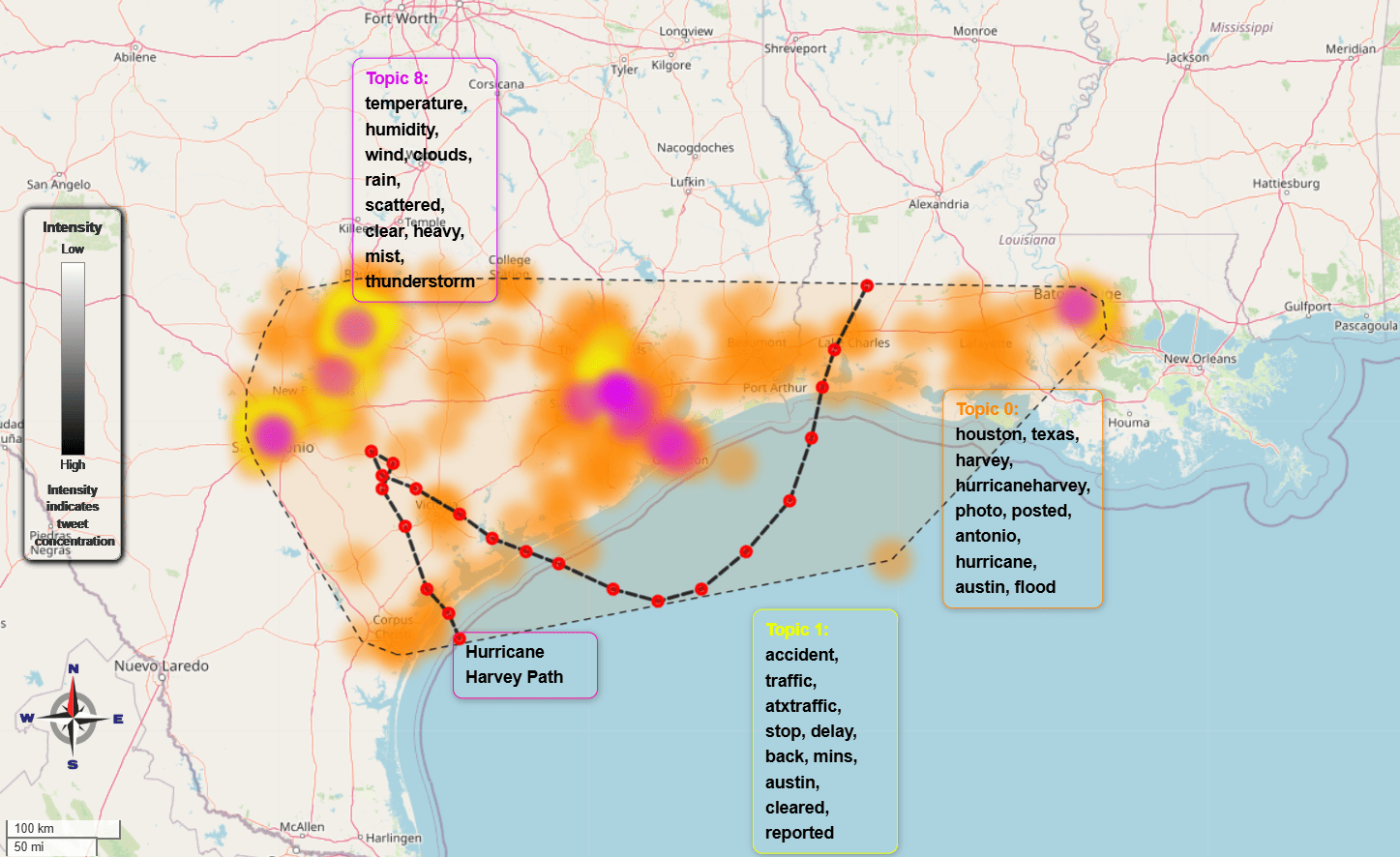}
    
    \vspace{0.1cm}
    
    \includegraphics[ width=11cm, frame]{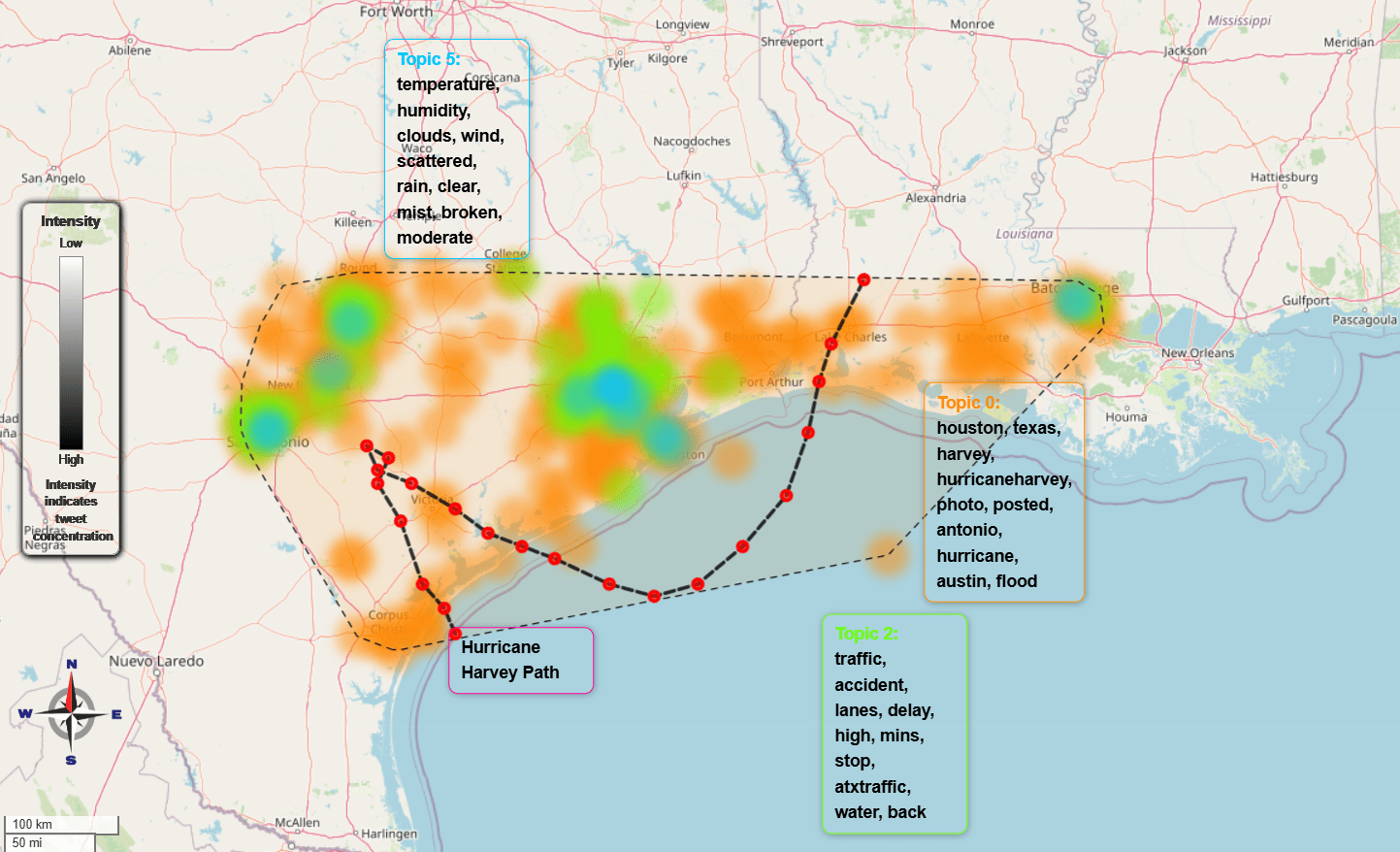}
    
    \vspace{0.4cm}
        
    \caption{Spatial distribution (KDE) of the Topic-words for a subset of the Hurricane Harvey dataset, applying MonoGraph  (top) and the MultiGraph  (bottom) models.}\label{fig:kdes}
\end{figure}

To gain qualitative insights into the latent structure of the discovered clusters, we analyzed both the semantic coherence and spatial distribution of topics generated by the MonoGraph and MultiGraph models applied to a subset of the dataset. The topic-word distributions, summarized in Fig.~\ref{fig:keywords}, reveal semantically coherent and lexically diverse clusters that reflect key aspects of disaster-related discourse, and  several topics capture geographically grounded and event-focused narratives. In particular, Topic~0 groups terms such as \textit{``houston,'' ``texas,'' ``harvey,'' ``hurricaneharvey,'' ``hurricane''} and \textit{``flood,''} reflecting a strong spatial anchoring of disaster-related conversations. As reflected in the coresponding spatial visualization in Fig.~\ref{fig:kdes}, the distribution of this topic closely follows the hurricane’s path, with dense tweet concentrations in Houston and along the Texas coastline. This alignment suggests that the models successfully identify localized, event-driven public responses directly associated with Hurricane Harvey’s impact zone. Another  distinct cluster is Topic~2 in MonoGraph and Topic~1 and in MultiGraph, centered on traffic-related concerns, including terms like \textit{``traffic,'' ``accident,'' ``mins,'' ``delay,''} and \textit{``stop.''} the spatial maps reveal concentrated tweet activity along major transport corridors in southeastern Texas, likely corresponding to road closures, congestion, and transportation disruptions resulting from flooding and evacuation efforts.

Other topics represent broader or more indirect engagement with the event. Topic~8 in MonoGraph and Topic~5 and in MultiGraph, captured meteorological observations, with keywords such as \textit{``temperature,'' ``humidity,'' ``clouds,'' ``rain,''} and \textit{``wind,''} are more evenly distributed across central Texas, including Austin and San Antonio. These spatial patterns suggest contributions from users outside the most severely affected areas, often in the form of commentary or weather updates rather than first-hand impact reports. Weather-related discourse is further reflected in Topics~2, 7, 5 and 9, in MonoGraph and Topics~3, 4, 9, 6 and 8 in MultiGraph, featuring overlapping terms such as \textit{``pressure,'' ``gust,'' ``humidity,'' ``wind,''} and \textit{``clouds.''} While this lexical overlap reduces topic distinctiveness, it underscores the recurring presence of weather-related signals across the corpus, a common characteristic in disaster-related social media data. This also illustrates the trade-off of fixing the number of topics ($k_{\text{means}} = 15$), which facilitates stability but can lead to partial redundancy. Finally, content-sharing and localized observational behavior emerge as a consistent pattern. For example, certain clusters contain terms such as \textit{``posted,'' ``photo,''} and place names like \textit{``meyerland,'' ``katy,''} and \textit{``downtown,''} suggesting the dissemination of real-time visual documentation and location-tagged updates by affected residents. This pattern reflects well-known crisis communication behavior on social media, where users share on-the-ground images to inform others and raise awareness.

\begin{figure}[t!]
    \centering
    \includegraphics[ width=11cm, frame]{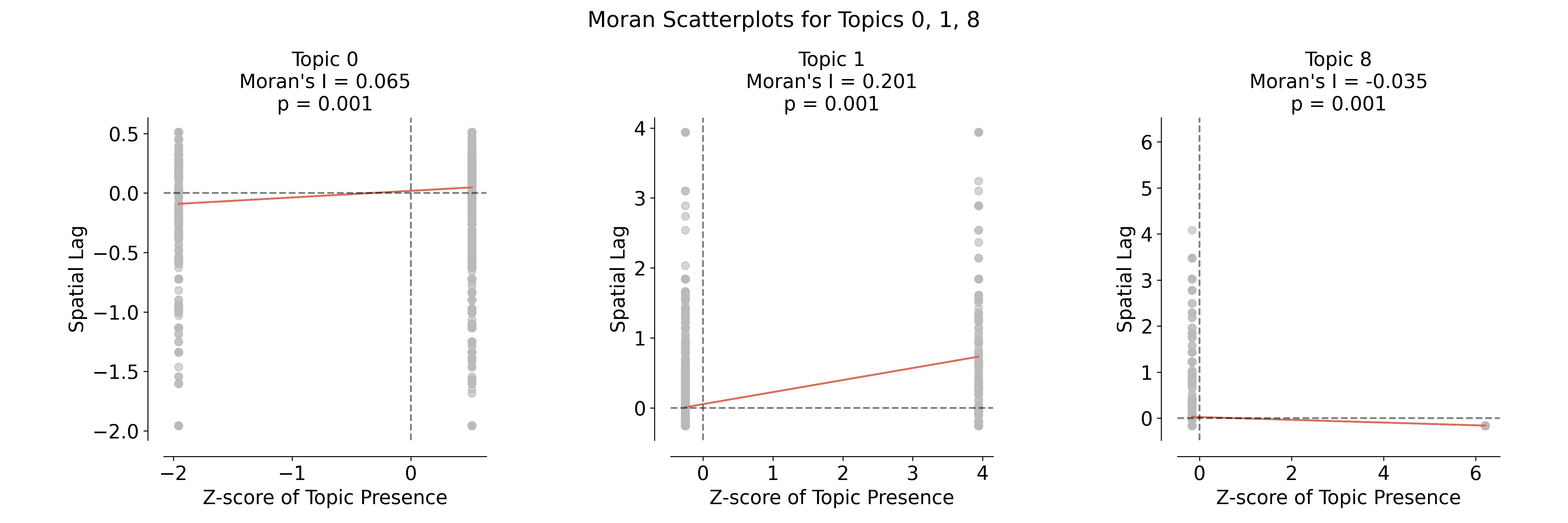}
    
    \vspace{0.1cm}
    
    \includegraphics[ width=11cm, frame]{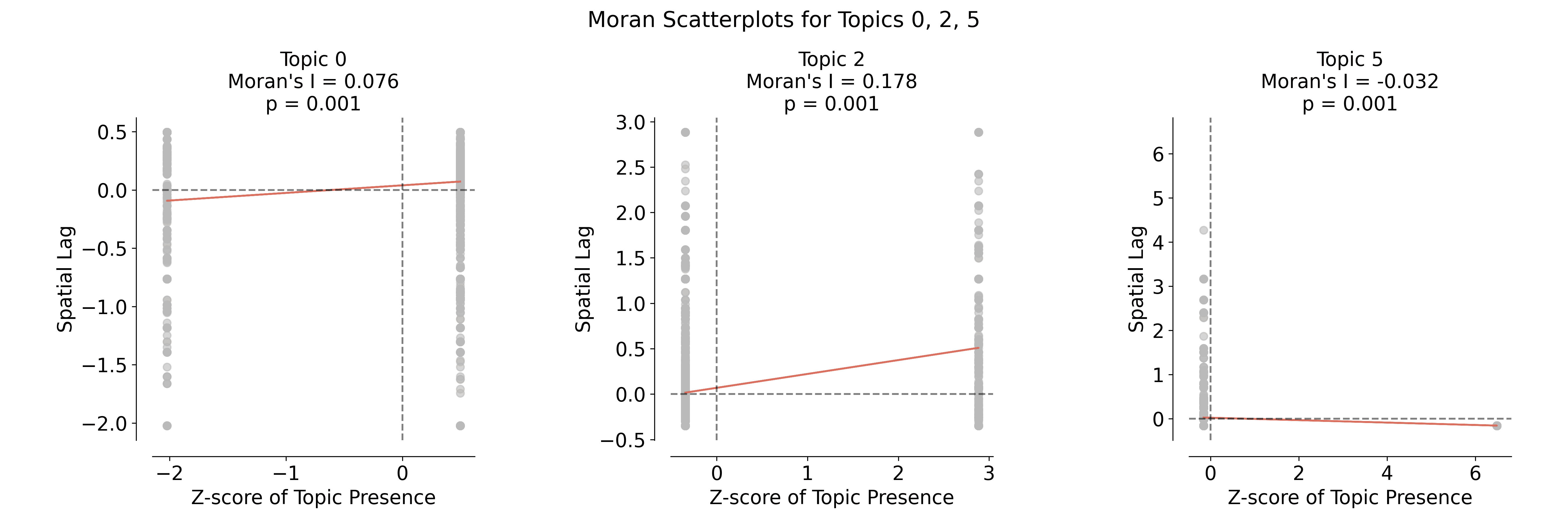}
    
    \vspace{0.4cm}
        
    \caption{Moran’s I plots for MonoGraph (top) and MultiGraph (bottom) Models.}\label{fig:morans}
\end{figure}

\subsection{Global Autocorrelation Interpretation}\label{5.2}

To explore the global distribution of tweet activity associated with different topics, we applied Moran's I analysis. This method is particularly beneficial as it helps to identify spatial autocorrelation, revealing whether tweet activity related to specific topics is clustered in certain geographic regions or distributed randomly around. The metric ranges from -1 (perfect dispersion) to +1 (perfect clustering), with 0 indicating randomness. A p-value $<$ 0.05 suggests the observed spatial pattern is statistically significant and unlikely due to chance. The related scatter plots and their associated statistics, demonstrated in Fig.~\ref{fig:morans} complement the spatial intensity patterns visualized in Fig.~\ref{fig:kdes}. In the MultiGraph model, Topic~0 shows a Moran’s I of 0.076 (p = 0.001), suggesting weak but statistically significant clustering. Tweets on this topic tend to occur more frequently near other tweets discussing the same theme than by chance, although the clustering is not very strong. The slight clustering might indicate localized interest or discussions related to this topic in specific regions. Topic~2 has a Moran’s I of 0.178 (p = 0.001), indicating weak but significant spatial clustering. The positive value suggests a tendency for tweets on this topic to be more closely located in certain regions. This could reflect a localized trend where users in particular areas are more engaged with or affected by the theme of the topic. Topic~5 has a Moran’s I of -0.032 (p = 0.001), a slightly negative Moran’s I indicating weak spatial dispersion, with tweets being more likely to occur near unrelated content. This suggests a dispersed pattern of topic-related tweets, though the effect is weak. The scattered distribution might indicate that this topic is more widespread, with no strong geographic concentration.

In the MonoGraph model, Topic~0 shows a Moran’s I of 0.065 (p = 0.001), which is similar to the MultiGraph model, showing weak clustering, though the clustering effect is slightly weaker than in the MultiGraph, with tweets on this topic being somewhat more scattered. This could imply a broader discussion, not concentrated in specific areas, though still more frequent in certain locations compared to random. Topic 1 has a Moran’s I of 0.201 (p = 0.001), indicating moderate clustering. Tweets on this topic show a clear tendency to be geographically concentrated in certain regions, which suggests a more localized theme. The higher Moran’s I value suggests that the topic is more likely to spark concentrated discussions within specific areas, potentially reflecting regional events or discussions. Topic 8 has a Moran’s I of -0.035 (p = 0.001), indicating a weakly dispersed distribution of tweets, showing no significant clustering and pointing to random spatial dispersion. The lack of clustering suggests that the topic has less geographic relevance or is more universally spread, with little concentration in specific regions.

Overall, most topics exhibit weak spatial clustering, as indicated by positive but modest Moran’s I values for topics 0, 1, and 2, reflecting slight regional concentration without distinct spatial boundaries. Topic 0, directly related to disasters, shows some geographic focus, while topics 1 and 2, indirectly related to disaster events, display weaker clustering. In the MonoGraph model, Topic 1 demonstrates stronger clustering (Moran’s I = 0.201), suggesting a more localized concentration. Topics 5 and 8 show weak spatial dispersion, with low or negative Moran’s I values, indicating a more even distribution across geographic regions and a lack of clear spatial patterns.

\begin{figure}[htbp]
    \centering
    \includegraphics[width=\textwidth]{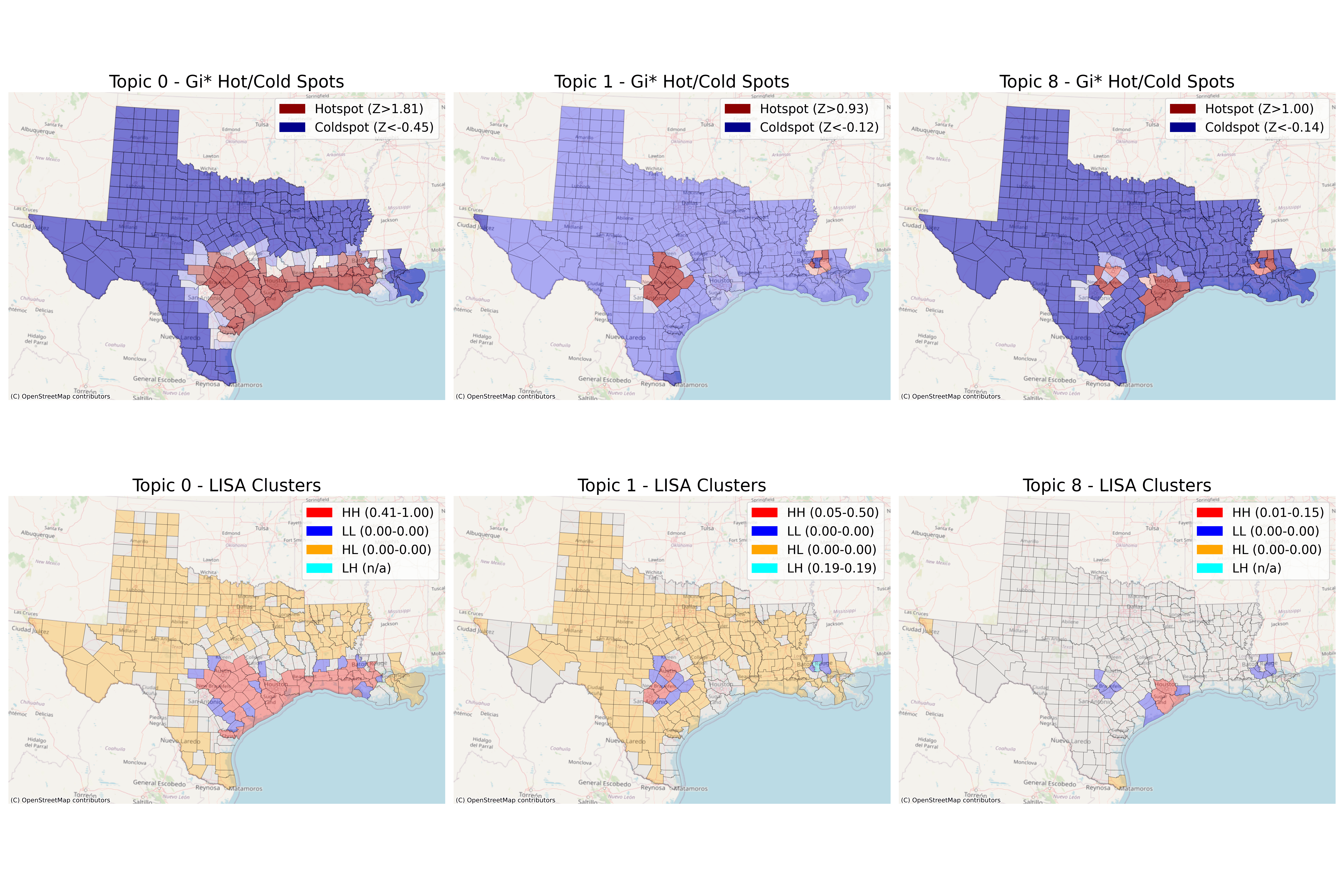}
    \caption{Spatial autocorrelation analysis (the Gi* maps (top) and LISA maps (bottom)) for the MonoGraph model.}
    \label{fig:SGLisa}
\end{figure}

\begin{figure}[htbp]
    \centering
\includegraphics[width=\textwidth]{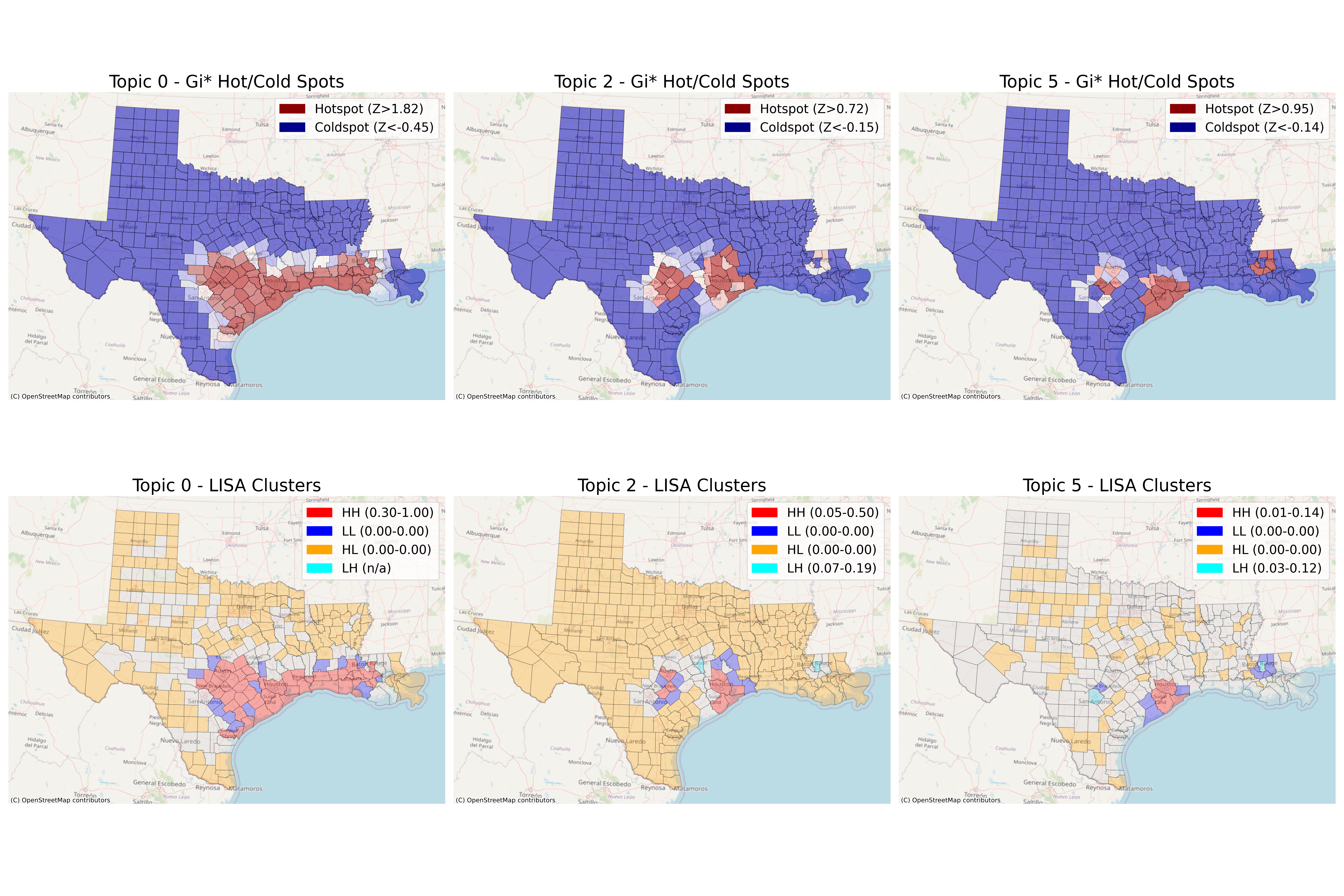}
    \caption{Spatial autocorrelation analysis (the Gi* maps (top) and LISA maps (bottom)) for the MultiGraph model.}
    \label{fig:MGLisa}
\end{figure}

\subsection{Spatial Autocorrelation Analysis}\label{sec5.2}

To further investigate the spatial (local) dynamics of identified topics, we applied the Getis-Ord Gi* statistic and Local Indicators of Spatial Association (LISA) to both the MonoGraph outputs. From a disaster management perspective, combining Gi* with LISA provides a powerful way to interpret social media–derived topics in both their intensity and spatial context. Gi* hotspots highlight areas of concentrated disaster-related activity, helping to quickly pinpoint zones of immediate impact for targeted relief. LISA adds a layer of context by classifying regions into high–high, low–low, high–low, and low–high clusters, showing not only where activity is high but also how it relates to surrounding areas. This combined approach enables planners to identify both core impact zones and peripheral but relevant areas, anticipate spillover effects, and spot isolated needs. In our results, the MultiGraph (Fig.~\ref{fig:MGLisa}) tends to produce more localized clusters, which may be better suited for tactical, on-the-ground coordination, while the MonoGraph model’s broader patterns (Fig.~\ref{fig:SGLisa}) can inform strategic, statewide planning. For Topic~0 in both models, representing core Hurricane Harvey discourse, the Gi* analysis revealed statistically significant hotspots spanning Houston, the Texas Gulf Coast, and inland counties along the storm’s path, with corresponding coldspots in northern and western Texas. LISA clustering confirmed persistent high–high (HH) clusters in these impact zones, indicating strong spatial concentration of disaster-related conversation. Such patterns can guide response agencies toward areas most actively engaged in reporting severe impacts, closely aligning with ground-truth disaster footprints.

As it can be seen in Figs.~\ref{fig:SGLisa} and \ref{fig:MGLisa}, traffic-related topics (MonoGraph Topic~2, MultiGraph Topic~1) produced more localized hotspots in and around metropolitan Houston and along major transport corridors. Gi* maps highlight narrowly bounded high-value clusters, while the LISA maps show HH concentrations surrounded by low–low (LL) areas, indicating geographically confined yet intense transportation disruptions. This suggests that mobility restoration efforts could be prioritized for specific corridors rather than across broad regions. Weather observation topics (MonoGraph Topic~8 MultiGraph Topic~5) exhibited scattered hotspots interspersed with wide coldspot zones. LISA classifications revealed fewer HH clusters but a notable presence of low–high (LH) outliers, counties with high weather-related discourse surrounded by low-activity neighbors. These may represent isolated monitoring stations, amateur weather observers, or unaffected communities contributing updates, which can complement official meteorological networks during disaster response.

\section{Discussion and Limitations}\label{sec6}

Across all datasets, both MonoGraph and MultiGraph consistently outperform baseline methods, achieving higher intra-cluster cohesion and lower inter-cluster similarity. Their relative strengths vary with event characteristics, reflecting sensitivity to spatial scale, discourse structure, and data quality. Fixing the number of clusters at $k_{\text{means}} = 15$ facilitates direct cross-model comparison and stabilizes evaluation, but also introduces constraints. In particular, a fixed $k_{\text{means}}$ may not reflect the true topical complexity of heterogeneous events. In some cases, semantically similar clusters emerge, especially for weather-related topics where recurring terms such as \textit{"pressure," "humidity"} and \textit{"wind"} are distributed across multiple clusters. This effect may be further amplified in datasets containing unfiltered automated or bot-generated content. Occasional empty or low-content clusters also appear in sparsely populated or noisy regions, likely due to limited message coverage or embedding collapse.

From a spatial perspective, event structure strongly shapes clustering behavior. MultiGraph produces compact, well-defined spatial clusters with fewer outliers in geographically concentrated events (e.g., Chile wildfires and the Napa earthquake), benefiting from modality-specific graph construction and cross-attention fusion. MonoGraph exhibits more robust separation in smaller or noisier areas of interest (e.g., Ahr Valley floods and Hurricane Harvey), where a unified representation better accommodates diffuse or overlapping discourse. As a result, MultiGraph tends to align clusters more closely with localized hazard impacts, while MonoGraph captures broader spatial narratives but may overestimate geographic extent.

Several limitations of the proposed framework naturally motivate directions for future research. Model performance depends on the quality and balance of the input modalities, which may vary across platforms and events and affect generalization. Temporal dynamics are not explicitly modeled, limiting the ability to capture topic evolution and cascading effects over time. While fixing $k_{\text{means}}$ enables controlled comparison across methods, adaptive or stability-driven strategies for determining the number of clusters, such as consensus-based or non-parametric approaches, represent an important extension. In addition, automated or bot-generated content is not explicitly filtered, which may introduce redundancy or noise in certain datasets. Evaluation currently relies on intrinsic topic quality metrics and spatial statistics; incorporating human-centered assessment, such as expert judgment of topic interpretability and relevance, or downstream task-based validation, would further strengthen claims of practical utility. Finally, the scalability of the MultiGraph architecture to real-time, high-volume data streams remains untested and may require distributed or streaming graph-processing solutions, alongside exploration of alternative graph construction strategies, additional modalities (e.g., sentiment or temporal signals), and explicit social network structures to enhance community-level insights.

\section{Conclusion}\label{sec7}

This work presented an unsupervised multimodal graph-based methodology for the semantic and spatial analysis of geo-social media data. By jointly modeling semantic embeddings and geographic coordinates through graph-based architectures, the approach addresses limitations of prior models that treat these modalities in isolation. Two architectural variants, MonoGraph and MultiGraph, were introduced to accommodate differing data characteristics and application needs. A composite loss function integrating contrastive, coherence, and alignment terms was key to learning robust, cluster-aware representations that preserve both local and global structure. Extensive evaluation on four real-world disaster datasets demonstrated consistent improvements over state-of-the-art baselines in topic coherence, spatial compactness, and overall topic quality. The MultiGraph variant, employing modality-specific encoding with attention-based fusion, achieved the strongest performance in most settings, while qualitative analysis of Hurricane Harvey (case study) confirmed the interpretability and geographic relevance of the discovered topics. The proposed methodology offers a scalable, principled approach to integrate symbolic and spatial signals for geo-social media analysis. Although the current implementation focuses on text and location, it can be extended to additional modalities such as sentiment, temporal dynamics, or imagery, enabling richer, more context-aware analyses in domains including disaster management, urban planning, and public health surveillance. By unifying multimodal semantic and spatial signals within an end-to-end learning framework, this work takes a step toward more holistic and interpretable models for understanding the complex information landscapes of geo-social media.

\bmhead{Acknowledgements}

This research was funded by the Austrian Research Promotion Agency (FFG) through the project \emph{MOSAIK} (grant number~926200) and by the European Union under the Horizon Europe Research and Innovation Actions programme (grant agreement No.~101093003, HORIZON-CL4-2022-DATA-01-01). Views and opinions expressed are, however, those of the author(s) only and do not necessarily reflect those of the European Union or the European Commission. Neither the European Union nor the European Commission can be held responsible for them.









\bibliography{references}

\end{document}